\documentclass[useAMS,usenatbib,usegraphicx]{mn2e}

\newcommand{\kms}{km s$^{-1}$}

\newcommand{\lam}{$\lambda$}

\newcommand{\civ}{\mbox{C\,{\sc iv}}}

\newcommand{\siiv}{\mbox{Si\,{\sc iv}}}

\newcommand{\nv}{\mbox{N\,{\sc v}}}

\newcommand{\pv}{\mbox{P\,{\sc v}}}

\title[Variability in quasar BAL outflows $-$ III]
{Variability in quasar broad absorption line outflows III. What happens on the shortest time-scales?}
\author[D. M. Capellupo et al.]{D. M. Capellupo$^{1,2}$
\thanks{E-mail:danielc@wise.tau.ac.il (DMC)}, F. Hamann$^{1}$, J. C. Shields$^{3}$, J. P. Halpern$^{4}$, and T.A. Barlow$^{5}$\\
$^{1}$Department of Astronomy, University of Florida, Gainesville, FL 32611-2055\\
$^{2}$School of Physics and Astronomy, Tel Aviv University, Tel Aviv 69978, Israel\\
$^{3}$Department of Physics \& Astronomy, Ohio University, Athens, OH 45701\\
$^{4}$Department of Astronomy, Columbia University, New York, NY 10027\\
$^{5}$Infrared Processing and Analysis Center, California Institute of Technology, Pasadena, CA 91125}
\begin{document}


\pagerange{\pageref{firstpage}--\pageref{lastpage}} \pubyear{2002}

\maketitle

\label{firstpage}

\begin{abstract}
Broad absorption lines (BALs) in quasar spectra are prominent signatures of high-velocity outflows, which might be present in all quasars and could be a major contributor to feedback to galaxy evolution. Studying the variability in these BALs allows us to further our understanding of the structure, evolution, and basic physical properties of the outflows. This is the third paper in a series on a monitoring programme of 24 luminous BAL quasars at redshifts 1.2 $< z <$ 2.9. We focus here on the time-scales of variability in \civ\ \lam1549 BALs in our full multi-epoch sample, which covers time-scales from 0.02$-$8.7 yr in the quasar rest-frame. Our sample contains up to 13 epochs of data per quasar, with an average of 7 epochs per quasar. We find that both the incidence and the amplitude of variability are greater across longer time-scales. Part of our monitoring programme specifically targeted half of these BAL quasars at rest-frame time-scales $\le$2 months. This revealed variability down to the shortest time-scales we probe (8$-$10 days). Observed variations in only portions of BAL troughs or in lines that are optically thick suggest that at least some of these changes are caused by clouds (or some type of outflow substructures) moving across our lines of sight. In this crossing cloud scenario, the variability times constrain both the crossing speeds and the absorber locations. Specific results also depend on the emission and absorption geometries. We consider a range of geometries and use Keplerian rotational speeds to derive a general relationship between the variability times, crossing speeds, and outflow locations. Typical variability times of order $\sim$1 year indicate crossing speeds of a few thousand km/s and radial distances near $\sim$1 pc from the central black hole. However,  the most rapid BAL changes occurring in 8-10 days require crossing speeds of 17 000 $-$ 84 000 \kms\ and radial distances of only 0.001$-$0.02 pc. These speeds are similar to or greater than the observed radial outflow speeds, and the inferred locations are within the nominal radius of the broad emission line region.

\end{abstract}


\section{Introduction}

High-velocity outflows are an integral part of the quasar system and likely contribute to feedback to the host galaxy. The material accreting onto the central supermassive black hole (SMBH) in quasar systems might release its angular momentum through these outflows. Moreover, these outflows might inject sufficient kinetic energy into the host galaxy to affect star formation, to aid in the `unveiling' of dust-enshrouded quasars, and to aide in the distribution of metal-rich gas to the intergalactic medium (\citealt{DiMatteo05}; \citealt{Moll07}).

Many properties of these outflows, including their location and three-dimensional structure, are still poorly understood. Current models of these outflows predict that they originate from the rotating accretion disc within the quasar system and are accelerated to high speeds via radiative and/or magneto-centrifugal forces (\citealt{Murray95}; \citealt{Proga00}; \citealt{Proga04}; \citealt{Everett05}; \citealt{Proga07}). In order to test these models and estimate the mass-loss rates and kinetic energy yields of these outflows, improved observational constraints are necessary. Understanding these properties will help to illuminate the role of quasar outflows in feedback to the host galaxy.

The most prominent signatures of these accretion disk outflows that are detected in quasar spectra are broad absorption lines (BALs). A quasar is classified as a BAL quasar if it has \civ\ \lam1549 absorption that reaches depths of least 10 per cent below the continuum over $>$2000 \kms.

This work is the third paper in a series on the variability in these BALs. We introduce our ongoing monitoring programme of 24 BAL quasars in \citeauthor{Capellupo11} (\citeyear{Capellupo11}, hereafter Paper 1). We begin with a sample of quasars from the BAL variability study in \citet{Barlow93}, and we have re-observed these quasars to obtain both a longer time baseline between observations and multiple epochs of data per object. In Paper 1, we directly compare \civ\ BAL variability properties between a ``short-term" interval of 0.35$-$0.75 yr\footnote{Throughout this paper, all time intervals are measured in the rest frame of the quasar.} and a ``long-term" interval of 3.8$-$7.7 yr. In particular, we report that 65 per cent of quasars exhibited BAL variability in the long-term, compared to 39 per cent in the short-term data. \citet{Gibson08} detected variability in 92 per cent (12/13) of quasars on similar time-scales as our ``long-term" data. This indicates that a majority of BAL quasars vary on multi-year time-scales. We also find a slight trend towards a larger change in strength in the long-term. These results are broadly consistent with the results of previous work on \civ\ BAL variability, i.e. \citet{Lundgren07} and \citeauthor{Gibson08} (\citeyear{Gibson08}; \citeyear{Gibson10}). The second paper in our BAL variability series, \citeauthor{Capellupo12} (\citeyear{Capellupo12}, hereafter Paper 2), includes all of the data obtained through 2009 March on our sample of 24 BAL quasars. We find that by adding more epochs of data, even at similar time-scales as the measurements in Paper 1, the fraction of quasars with variable BAL absorption increases.

Paper 2 also discusses in detail the two prevailing scenarios that are likely causing the observed BAL variability, i.e., a change in ionization in the outflowing gas and outflow clouds, or substructures, moving across our line of sight. Using the results of Paper 2, along with the results of previous work, we can not rule out either scenario. Instead, Paper 2 concludes that the actual situation is likely a complex amalgam of both scenarios. Understanding the cause(s) of variability is necessary for using variability to constrain properties of the flows. In the moving-cloud scenario, the time-scales of variability can help constrain the sizes, speeds, and locations of clouds. For example, in Paper 1 we estimate that the distance of the outflowing gas is roughly within 6 pc of the central SMBH for those BALs that varied on time-scales of $\sim$0.5 yr. Variability on shorter time-scales would place the outflow at even smaller distances, based on nominally shorter crossing times for clouds moving across our line of sight (\citealt{Hamann08}; Paper 1). In the changing ionization scenario, variability on shorter time-scales require high densities due to the shorter recombination times, which would also indicate that the outflows are located near the emission source \citep{Hamann97}.

In the current work, we include all the data from our BAL monitoring programme. While earlier work on BAL variability have data on time-scales $\le$1 yr (\citealt{Barlow93}; \citealt{Lundgren07}; \citealt{Gibson10}; Papers 1 and 2), information on variability on time-scales $<$1$-$2 months is very limited. Therefore, we extended our monitoring programme to specifically probe the shortest time-scales, down to $\sim$1 week in the quasar rest-frame. As mentioned above, variability on short time-scales, or lack thereof, will help constrain the location of these outflows, which is crucial for understanding the physics of the flows. Furthermore, no matter what causes the variability, variability on the shortest time-scales would require extreme circumstances, such as very rapid continuum flux changes, very rapid crossing speeds, and/or small outflow structures.

Besides examining the shortest time-scales, we also use the multi-epoch nature of our dataset, which spans a wide range of time intervals ($\Delta$$t$) from 0.02 yr (7 days) to 8.7 yr, to investigate the incidence of \civ\ BAL variability as a function of $\Delta$$t$. We look for any drop-off in the incidence of variability at the shortest time-scales, and at longer, multi-year time-scales, we investigate whether the variability fraction continues to increase as the time-scale increases, as in Paper 1.

We give an overview of the BAL monitoring programme in Section \ref{ch4.obs} and describe our analysis in Section \ref{ch4.meas}. In Section \ref{ch4.amp}, we investigate how changes in the absorption strength of BALs correlate with time-scale. We highlight specific cases of BAL variability on the shortest time-scales we probe in Section \ref{ch4.short}, and we examine trends in \civ\ BAL variability with time-scale in Section \ref{ch4.time}. In Section \ref{results}, we summarize the main results of this series of papers on BAL variability. We then discuss, in Section 5, potential causes of the observed BAL variability and the implications of the results of this work in terms of physical properties of the outflowing gas.

\section{Data and Analysis}
\label{ch4.data}

\subsection{Observations and quasar sample}
\label{ch4.obs}

This paper continues our analysis of the same 24 BAL quasars studied in Papers 1 and 2 (see table 1 in Papers 1 and 2 for more information on the sample). This sample was defined in \citet{Barlow93}, which contains data from the Lick Observatory 3-m Shane Telescope, using the Kast spectrograph. We include data from this study that have a resolution of $R\equiv\lambda/\Delta\lambda\approx 1300$ (230 \kms), or $R \approx$ 600 (530 \kms) if there is no higher resolution data available for that epoch. Either resolution is sufficient for studying the variability in BALs since they are defined to have a width of at least 2000 \kms.

We have been monitoring this sample of BAL quasars using the MDM Observatory 2.4-m Hiltner telescope with the CCDS spectrograph, with a resolution of $R \approx$ 1200 (250 \kms). We also include Sloan Digital Sky Survey (SDSS) Data Release 6 spectra, which have a resolution of $R\approx 2000$ (150 \kms), when available \citep{Adelman08}. We include in this work all of the observations described in Papers 1 and 2, and more details on the telescope and spectrograph setups can be found therein.

For the current work, we supplement the data from Papers 1 and 2 with additional observations of a subsample of our BAL quasars, covering rest-frame time intervals of $\Delta$$t$ $\sim$1 week to 1 month, during the first half of 2010. We chose objects with a right ascension of 08$-$15 hrs so that we could observe them from January through May. This restricts our sample to 17 out of 24 quasars, and out of these 17, we were able to observe 13 of the quasars 2 to 4 times each. We tended to give preference to quasars that were known to have variable \civ\ BALs in order to achieve our goal of testing for small variability time-scales in a reasonable amount of telescope time. Therefore, in the analysis below, we might be slightly biased towards higher variability fractions, especially at the shorter time-scales. We address this bias below, in Section \ref{ch4.time}, and find that the bias is minimal.

Most of these observations were taken at the MDM 2.4-m in 2010 January (2010.04), February (2010.13), March (2010.20), and May (2010.35). We also have some observations taken at the KPNO 2.1-m in 2010 February (2010.11). We used the GoldCam spectrograph with a 600 groove per mm grating in first order and a 2 arcsec slit, providing a spectral resolution of $R \approx$ 1100 (275 \kms). The observed wavelength coverage is $\sim$3600 to 6100 \AA\ for all the observations, which covers the full \siiv\ and \civ\ absorption regions for the quasars observed. The co-added exposure times were 2.5$-$3 h per source. The typical noise level for our observations are shown in fig. 1 of both Papers 1 and 2.

\begin{table*}
  \begin{minipage}{100mm}
    \caption{Full BAL Quasar Dataset and Shortest Time-Scale Data}
    \begin{tabular}{ccccccc}
\hline
Name & $z_{em}$ & BI & No. of & $\Delta$$t$ & No. of Meas. & $\Delta$$t$ $<$ 72 \\
 & & & Obs. & (yrs) & $<$72 days & (days) \\
\hline
    0019+0107		& 2.13 & 2290		& 7   & 0.08$-$5.79	& 2	& 28$-$30 \\
    0043+0048   	& 2.14 & 4330  		& 5   & 0.35$-$6.13	& ...	& ... \\
    0119+0310   	& 2.09 & 6070  		& 3   & 0.65$-$5.57	& ...	& ... \\
    0146+0142   	& 2.91 & 5780  		& 4   & 0.52$-$5.15	& ...	& ... \\
    0226$-$1024 	& 2.25 & 7770  		& 2   & 4.66 		& ...	& ... \\
    0302+1705   	& 2.89 & 0       		& 3   & 0.27$-$4.42	& ...	& ... \\
    0842+3431   	& 2.15 & 4430  		& 13 & 0.03$-$6.51	& 9	& 10$-$64 \\
    0846+1540   	& 2.93 & 0       		& 8   & 0.04$-$4.93	& 4	& 14$-$65 \\
    0903+1734   	& 2.77 & 10700		& 7   & 0.04$-$5.38	& 2	& 15$-$32 \\
    0932+5006   	& 1.93 & 7920  		& 13 & 0.02$-$7.37	& 9	& 9$-$39 \\
    0946+3009   	& 1.22 & 5550  		& 9   & 0.03$-$8.67	& 7	& 12$-$51 \\
    0957$-$0535 	& 1.81 & 2670  		& 8   & 0.02$-$6.93	& 7	& 9$-$42 \\
    1011+0906   	& 2.27 & 6100  		& 10 & 0.03$-$6.55	& 5	& 10$-$37 \\
    1232+1325   	& 2.36 & 11000		& 3   & 0.35$-$5.93	& ...	& ... \\
    1246$-$0542 	& 2.24 & 4810  		& 7   & 0.02$-$6.52	& 3	& 8$-$25 \\
    1303+3048   	& 1.77 & 1390  		& 9   & 0.03$-$6.51	& 7	& 9$-$41 \\
    1309$-$0536 	& 2.22 & 4690  		& 7   & 0.02$-$6.54	& 3	& 8$-$25 \\
    1331$-$0108 	& 1.88 & 10400 	& 7   & 0.02$-$6.64	& 1	& 9 \\
    1336+1335   	& 2.45 & 7120   	& 4   & 0.07$-$5.79	& 1	& 27 \\
    1413+1143   	& 2.56 & 6810  		& 7   & 0.03$-$5.89	& 1	& 9 \\
    1423+5000   	& 2.25 & 3060  		& 8   & 0.02$-$6.49	& 3	& 8$-$25 \\
    1435+5005   	& 1.59 & 11500 	& 7   & 0.03$-$8.15	& 3	& 10$-$31 \\
    1524+5147   	& 2.88 & 1810  		& 9   & 0.02$-$5.39	& 3	& 7$-$32 \\
    2225$-$0534 	& 1.98 & 7920  		& 3   & 0.27$-$0.73	& ...	& ... \\
\hline
    \end{tabular}
    \label{ch4.table}
  \end{minipage}
\end{table*}

We incorporated this new data into our existing dataset to create one large dataset for our sample of 24 BAL quasars, covering time-scales, $\Delta$$t$, of 0.02 to 8.7 yr. Table \ref{ch4.table} summarizes the full dataset from our BAL monitoring programme, including the emission redshift, $z_{em}$,\footnote{The values of $z_{em}$ were obtained from the NASA/IPAC Extragalactic Database (NED), which is operated by the Jet Propulsion Laboratory, California Institute of Technology, under contract with the National Aeronautics and Space Administration.} and the `balnicity index' (BI) for each object (as calculated in Paper 1). The fourth column gives the total number of observations, and the fifth column gives the range in $\Delta$$t$ values for all the measurements of each object, where one measurement is a pair of observations. The addition of the most recent data described above to the existing sample gives a total of 17 BAL quasars with measurements that cover time-scales $<$0.20 yr (72 days). The sixth column lists the number of measurements at these short time-scales for these 17 quasars, and the final column gives the range in $\Delta$$t$ values, in days, for these measurements.

\subsection{Measuring BALs and their variability}
\label{ch4.meas}

This paper focuses on variability in \civ\ BALs over different time-scales. In order to measure variability, we first adopt the velocity intervals over which \civ\ BAL absorption occurs in each quasar defined in Paper 1. We used the definition of BI as a guide for defining these regions, i.e. they must contain contiguous absorption that reaches $\geq$10 per cent below the continuum across at least 2000 \kms.

In order to search for variability between two epochs, we used a simple vertical scaling, as described in more detail in Papers 1 and 2, to match all the spectra to the fiducial Lick spectrum defined in Paper 1. We consider all available unabsorbed spectral regions when scaling the spectra, and we focus especially on the spectral regions that immediately border the absorption lines of interest. In some cases, a simple scaling does not produce a good match, and there were disparities in the overall spectral shape between the comparison spectra. For these cases, we typically fit a linear function, or rarely, a quadratic function, to the ratio of the two spectra across regions that avoid the BALs and then multiplied this function by the MDM or KPNO spectrum to match the fiducial Lick data.

For this work, we use precisely the same criteria as described in Papers 1 and 2 for determining whether a quasar has a variable BAL. This process includes visually inspecting each pair of spectra for each quasar. For a pair of observations to qualify as an incidence of variability, the varying region(s) must meet two thresholds. First, the candidate varying region must be at least 1200 \kms\ in width. Second, we use the average flux and associated error within that velocity interval to place an error on the flux differences between the two spectra (see Equation 1 in Paper 2). The flux difference in this interval must be at least 4$\sigma$ to be included as a varying region.

In general, we take a conservative approach and omit a small number of ambiguous cases that meet this threshold. Our goal here is to avoid over-counting the occurrences of variability. Photon statistics alone are not sufficient for defining real variability because flux calibrations, a poorly constrained continuum placement, and underlying emission-line variability can all add additional uncertainty not captured by photon statistics. However, on the shortest time-scales, we find that the amplitudes and velocity intervals of variability both tend to be small (Paper 1; \citealt{Gibson08}; and, Section 3.1 below). We therefore consider two categories of short-term variability in our analysis, namely, 1) ``secure" cases that meet all of our criteria above, and 2) ``tentative" cases, which are likely real variations but are excluded by our conservative approach. We discuss several examples of secure and tentative detections of variability in Section \ref{ch4.short} below. We also looked for very short-term changes by comparing exposures taken on different nights within the same observing run. The constraints are poor due to lower signal-to-noise in individual exposures, but we checked for such changes within the 2010 data and found no variability.

In order to measure the absorption strength of individual troughs, we first adopt the power-law continuum fits defined in Papers 1 and 2. A power-law continuum is fit to one fiducial Lick spectrum per object. Since the broad emission lines (BELs) can also vary, we only apply the power-law fit to the other epochs included here, and we do not fit the emission lines. Any measurements of absorption strength in this paper, as in Section \ref{ch4.amp}, are performed at velocities that are mostly unaffected by the BELs, so fitting the BELs for each epoch is not necessary.

As in Papers 1 and 2, we first calculate the absorption strength, $A$, which is the fraction of the normalized continuum flux removed by absorption (0 $\le A \le$ 1). For each measurement of variability, we then calculate the average $A$ for each epoch within the velocity interval that varied and determine the change in strength, $\Delta$$A$, between the two epochs. We do not divide the variable intervals into bins as we do in Papers 1 and 2; instead, we calculate the average $\Delta$$A$ over the entire velocity interval that varied. In Papers 1 and 2, we find that variability tends to occur in just portions of troughs, and sometimes in very narrow portions (also \citealt{Gibson08}). These measurements of $A$ and $\Delta$$A$ give a direct measurement of the strength of the lines, and the change in strength, in only the portions that varied. Equivalent width (EW) measurements of a wide trough are less sensitive to strength changes in a narrow portion of the trough.

\section[]{Results}
\label{ch4.results}

\subsection{Amplitude of Variability}
\label{ch4.amp}

\begin{figure}
  \includegraphics[width=84mm]{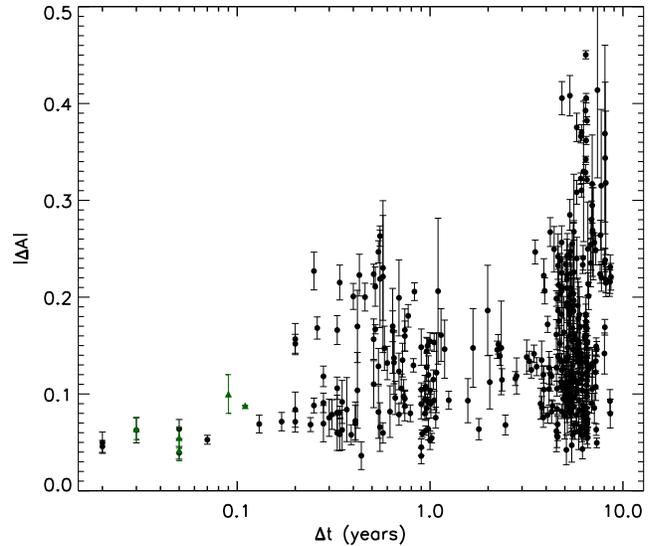}
  \caption{The change in absorption strength, $\Delta$$A$, for variable portions of absorption
    troughs versus the time interval between observations, $\Delta$$t$. The green triangles at
    shorter time-scales represent measurements of $\Delta$$A$ for the tentative cases of variability.}
  \label{dAvdt}
\end{figure}

In Paper 1, we directly compared the \civ\ BAL variability between a ``short-term" time interval of 0.35$-$0.75 yr and a ``long-term" interval of 3.8$-$7.7 yr, and the change in strength was generally slightly higher in the ``long-term" data. To further investigate whether there is truly a correlation between the amplitude of strength changes and the variability time-scale, we plot $\Delta$$A$ versus $\Delta$$t$ (Fig. \ref{dAvdt}). In order to avoid the added uncertainty of overlapping emission lines, we calculate $\Delta$$A$ only for absorption between $-$27 000 and $-$8 000 \kms\ (Paper 1). We find an increasing upper envelope of values of $\Delta$$A$ with increasing $\Delta$$t$. This is consistent with the findings of \citet{Gibson08}, who show that the envelope of values of $\Delta$EW expands as $\Delta$$t$ increases. We only plot variable cases in Fig. 1, as well as measurements of the tentative cases of variability (green triangles; Section \ref{ch4.meas}). On the shortest time-scales that we are probing in this paper, with $\Delta$$t$ $<$ 0.20, the $\Delta$$A$ values are all below $\sim$0.1. On multi-year time-scales, the values of $\Delta$$A$ reach as high as $\sim$0.46.

\subsection{Examples of Variability on the Shortest Time-Scales}
\label{ch4.short}

As mentioned in Section \ref{ch4.data}, we obtained new data specifically to augment the temporal sampling of our dataset at time-scales $<$0.20 yr ($<\sim$72 days) in the quasar rest-frame. We specifically look at this range of time-scales because it corresponds to the four shortest time-scale bins in Fig. 9 below, where we examine trends with the BAL variability time-scale. Out of the 17 quasars for which we have data on these short time-scales, only 2 quasars exhibited \civ\ BAL variability (12 per cent). If we include the tentative cases of variability, then 5 out of these 17 quasars varied (29 per cent). For comparison, even with the tentative cases included, this fraction is still lower than the incidence of variability in our ``short-term" sub-sample from Paper 1, where 39 per cent of the quasars varied over time intervals of 0.35$-$0.75 yr.

In this section, we first highlight the only two quasars with secure detections of \civ\ BAL variability on the shortest time-scales in our data set ($<$0.20 yr). We then describe examples of cases that were classified as tentative detections of variability on these short time-scales. We emphasize that all of the cases of variability, and tentative variability, discussed here meet the threshold defined in Paper 1 and used throughout this paper, i.e. the variability must occur over a region at least 1200 \kms\ wide and the flux difference between the two spectra must be at least 4$\sigma$ (see Section \ref{ch4.meas}). We present these individual cases in order of increasing time-scale.

\subsubsection{1246$-$0542: Secure variability over 8 Days}
\label{ch4.1246}

\begin{figure*}
  \includegraphics[width=100mm, angle=90]{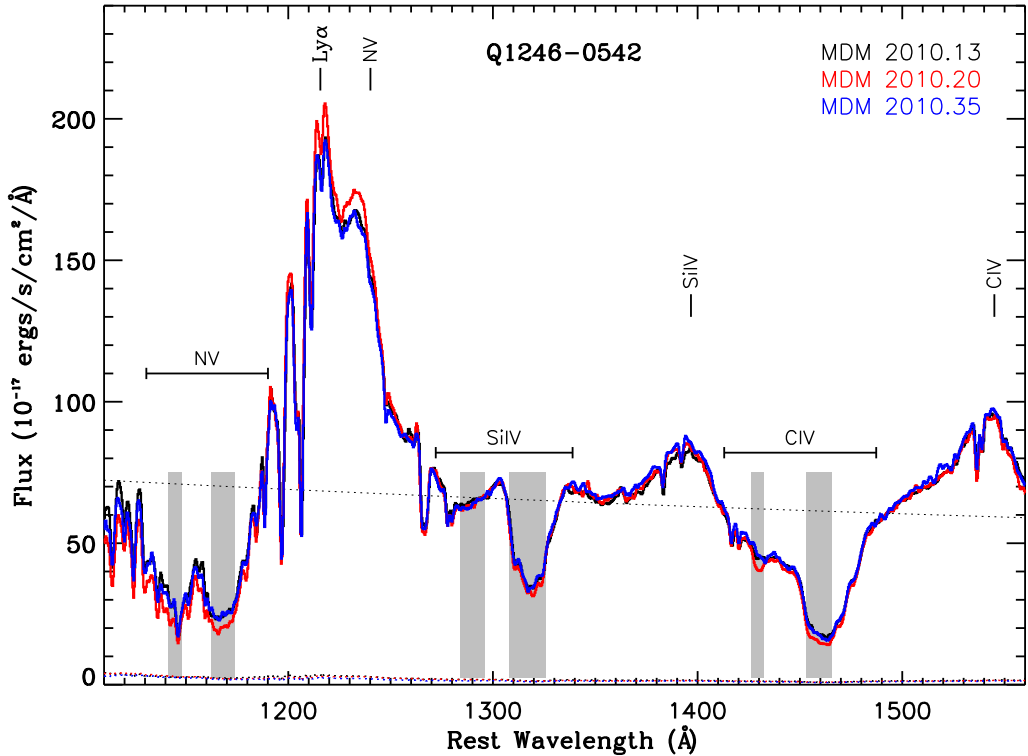}
  \caption{The full spectra of 1246$-$0542, showing 3 epochs covering a $\Delta$$t$ of 
    0.02$-$0.07 yr. The emission lines are identified by vertical bars. The rightmost horizontal bar
    marks the \civ\ BAL absorption, and the shaded regions below it mark the intervals of variability.
    The other horizontal bars and shaded regions mark the corresponding velocities in \nv\ and \siiv.
    We use binomial smoothing to improve the presentation of these and all of the other spectra
    displayed in this paper. The black dashed curve shows the power-law fit to the continuum, and
    the formal 1$\sigma$ errors for each spectrum are shown near the bottom.}
  \label{sp1246}
\end{figure*}

\begin{figure}
  \includegraphics[width=84mm]{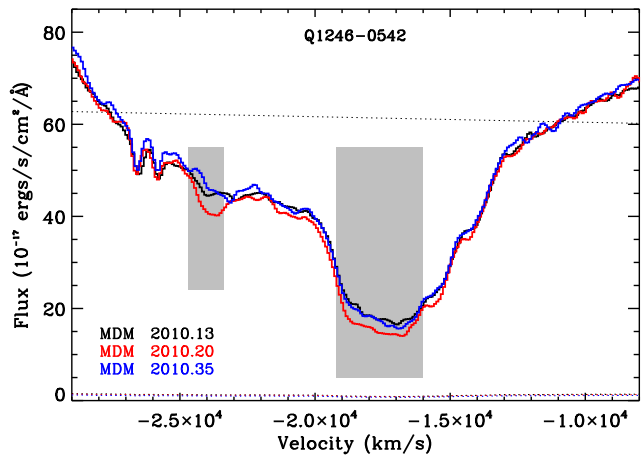}
  \caption{Spectra of just the \civ\ BAL in 1246$-$0542 for the same 3 epochs shown in Fig.
    \ref{sp1246}. The shaded regions mark the intervals of variability. As in Fig. \ref{sp1246}, the
    black dashed curve shows the power-law fit to the continuum, and the formal 1$\sigma$ errors
    for each spectrum are shown near the bottom.}
  \label{sp1246z}
\end{figure}

The quasar 1246$-$0542 is one of just two quasars in our sample for which we have a secure detection of variability over a time-scale $<$0.20 yr. Our dataset contains measurements down to $\Delta$$t$ of 0.02 yr ($\sim$7$-$8 days), and we found variability down to this time-scale in 1246$-$0542. We present this measurement in Fig. \ref{sp1246}, showing the full spectrum for the 2010.13 and 2010.20 observations, and in Fig. \ref{sp1246z}, showing just the \civ\ BAL. As indicated by shaded bars in Figs \ref{sp1246} and \ref{sp1246z}, there are two separate velocity intervals of variability between the 2010.13 (black curve) and 2010.20 (red curve) observations. The two velocity intervals are centered at $-$24 050 \kms and $-$17 200 \kms, with widths of 1300 \kms\ and 2600 \kms, respectively. The flux differences in these two intervals are 8.0$\sigma$ and 13$\sigma$.

The rightmost shaded bars in Fig. \ref{sp1246} mark the variable intervals in \civ, and they correspond to the shaded intervals in Fig. \ref{sp1246z}. We shifted these intervals to the corresponding velocities in \nv, marked by the leftmost shaded bars, and \siiv. The doublet separations for \nv\ and \siiv\ are 960 and 1930 \kms, respectively, which are both wider than the doublet separation for \civ\ (500 \kms). When shifting the shaded intervals to \nv\ and \siiv, we therefore widened the intervals based on the greater doublet separations in these lines. The identification of \nv\ absorption, and any potential variability, is complicated by Ly$\alpha$ forest absorption, which appears throughout the entire spectrum blueward of the Ly$\alpha$ \lam1216 emission line. There is also a slight mismatch between the two spectra near the limit of the wavelength coverage. However, we use the velocities of \civ\ absorption to identify where the \nv\ absorption should be located, and the velocities of apparent variability in \nv, in the redmost velocity interval especially, match the velocities at which \civ\ appears to vary. Corresponding variability in \nv\ supports the detection of variability in \civ. We also note that despite the slight mismatch at the limit of the wavelength coverage, the two spectra match well along the remainder of the spectral coverage and, in particular, in the spectral regions that immediately border the \civ\ absorption. We do not find any corresponding changes in the \siiv\ BAL. In Paper 2, we found just one tentative case where \civ\ variability was not accompanied by \siiv\ variability at the same velocity.

Figs \ref{sp1246} and \ref{sp1246z} also show that the BAL in 1246$-$0542 again varied between 2010.20 and 2010.35 ($\Delta$$t$ $\sim$ 0.05 yr, or 17 days), returning to its previous state from 25 days earlier. There is no difference in the BAL strength between the earlier 2010.13 spectrum (black curve) and the 2010.35 spectrum (blue curve). We also have measurements of this quasar at other epochs with variability in the same velocity intervals within which we detect variability in Figs \ref{sp1246} and \ref{sp1246z}. In particular, in Papers 1 and 2, we find variability, with a flux difference of 6$\sigma$, between the two ``long-term" epochs, 1992.19 and 2008.35, at the same velocities as the bluemost interval here (see fig. 1 in Papers 1 and 2). These ``long-term" epochs are two different observations that were taken at earlier epochs than the observations shown here in Figs \ref{sp1246} and \ref{sp1246z}. This gives further confirmation that the detection of variability over 8 days between 2010.13 and 2010.20 is secure and not a result of a poor flux calibration in one of the spectra or a poor match between the unabsorbed regions in the two spectra.

\subsubsection{0842+3431: Secure variability over 10 Days}
\label{ch4.0842}

\begin{figure*}
  \includegraphics[width=135mm]{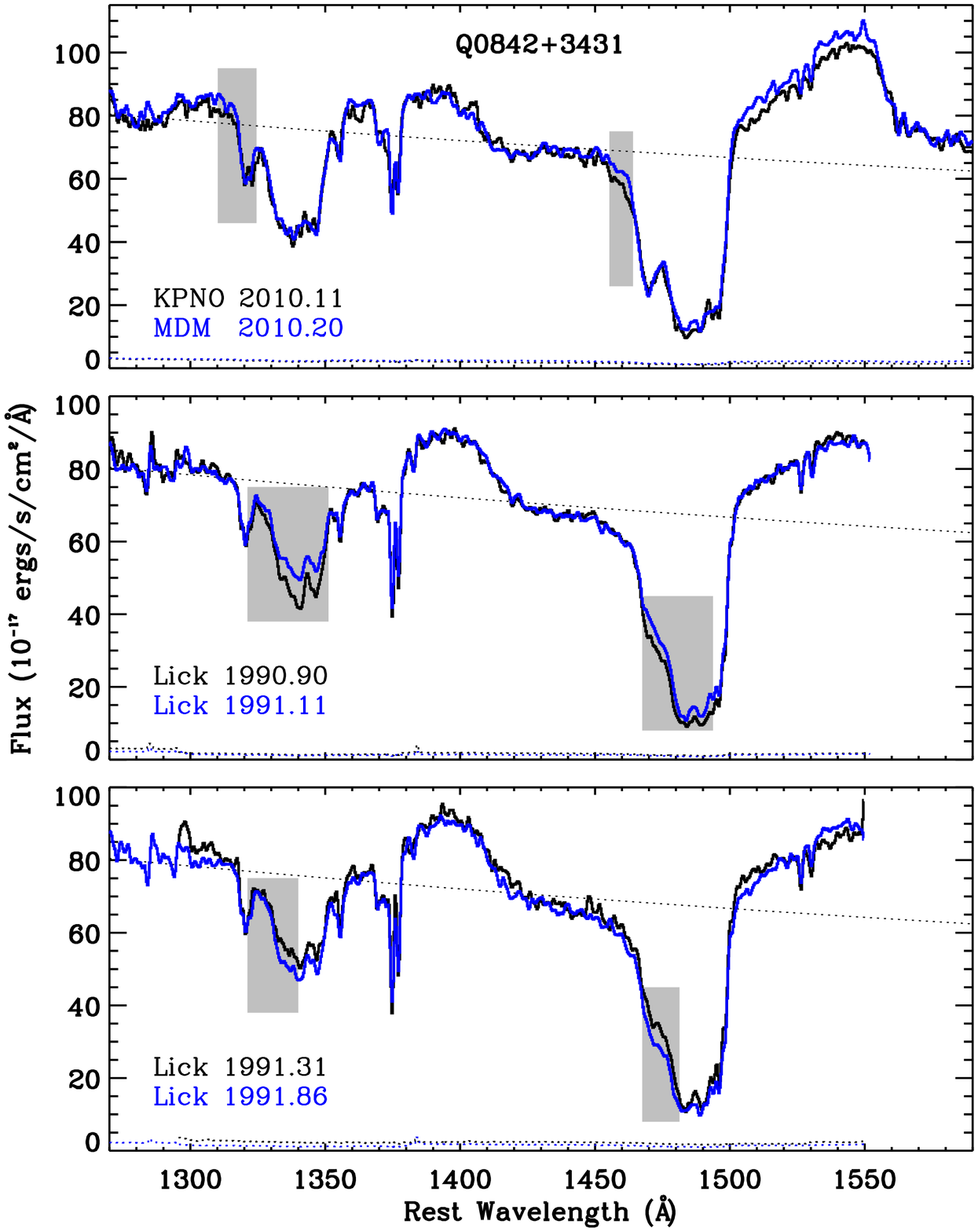}
  \caption{Spectra of 0842+3431. Each panel shows a measurement of variability over a time-scale
    of $<$0.20 yr, with the top panel showing two epochs separated by just 0.03 yr (10 days). The
    rightmost shaded regions mark the intervals of \civ\ variability for each measurement, and the
    leftmost shaded regions mark the corresponding velocities in \siiv.}
  \label{sp0842}
\end{figure*}

\begin{figure}
  \includegraphics[width=84mm]{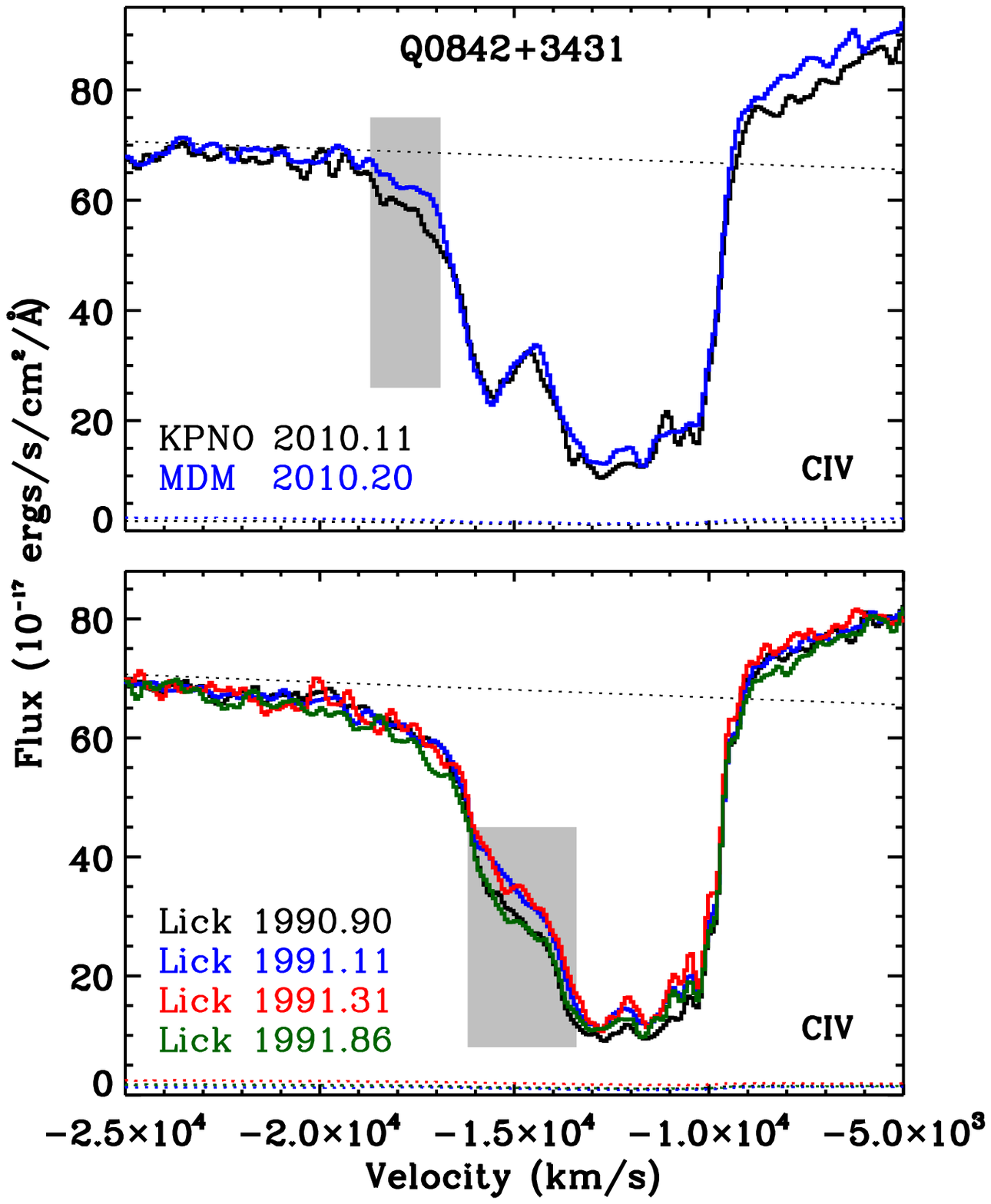}
  \caption{Spectra of the \civ\ BAL in 0842+3431. The top panel shows the same two epochs from
    the top panel of Fig. \ref{sp0842}, and the bottom panel shows all four Lick epochs from the
    bottom two panels of Fig. \ref{sp0842}. The shaded regions mark the extent of the \civ\ variability.}
  \label{sp0842z}
\end{figure}

The quasar 0842+3431 is the only quasar besides 1246$-$0542 with a secure detection of variability at $\Delta$$t$ $<$ 0.20 yr. We detect variability over 0.03 yr (10 days) between the 2010.11 and 2010.20 observations of 0842+3431. These spectra are shown in the top panel of both Figs \ref{sp0842} and \ref{sp0842z}, and the variable interval, marked by a shaded bar, is centered at $-$17 800 \kms\ and is 1800 \kms\ wide. The flux difference between the two epochs is 8.5$\sigma$.

This variable interval is on the blue side of the trough at a high enough velocity that this detection of variability is not affected by any variability in the \civ\ BEL. The spectra also match very well on either side of the shaded interval. Furthermore, we mark the corresponding velocities in \siiv\ with a shaded bar, and there is potential variability there.

The middle panel of Fig. \ref{sp0842} shows two Lick spectra, 1990.90 and 1991.11, which are separated by 24 days, and the bottom panel shows the 1991.31 and 1991.86 spectra, which are separated by 64 days. In both of these measurements, there is variability in \siiv\ at the same velocities as the variability in \civ. Between 1990.90 and 1991.11, the variability is stronger towards the blue and red ends of the interval, but much weaker in the middle of the interval. The variability in \siiv, however, is most significant in the middle of the shaded interval. All four spectra from 1990.90 to 1991.86 are shown in the bottom panel of Fig. \ref{sp0842z} (a full discussion of the multi-epoch nature of the variability in 0842+3431 can be found in Paper 2). The clear low-amplitude variability on 24 and 64-day time-scales in the Lick data support our determination of similar variability (albeit at different velocities) over 10 days between 2010.11 and 2010.20.

\subsubsection{1011+0906: Tentative variability over 17 days}
\label{ch4.1011}

\begin{figure*}
  \includegraphics[width=100mm, angle=90]{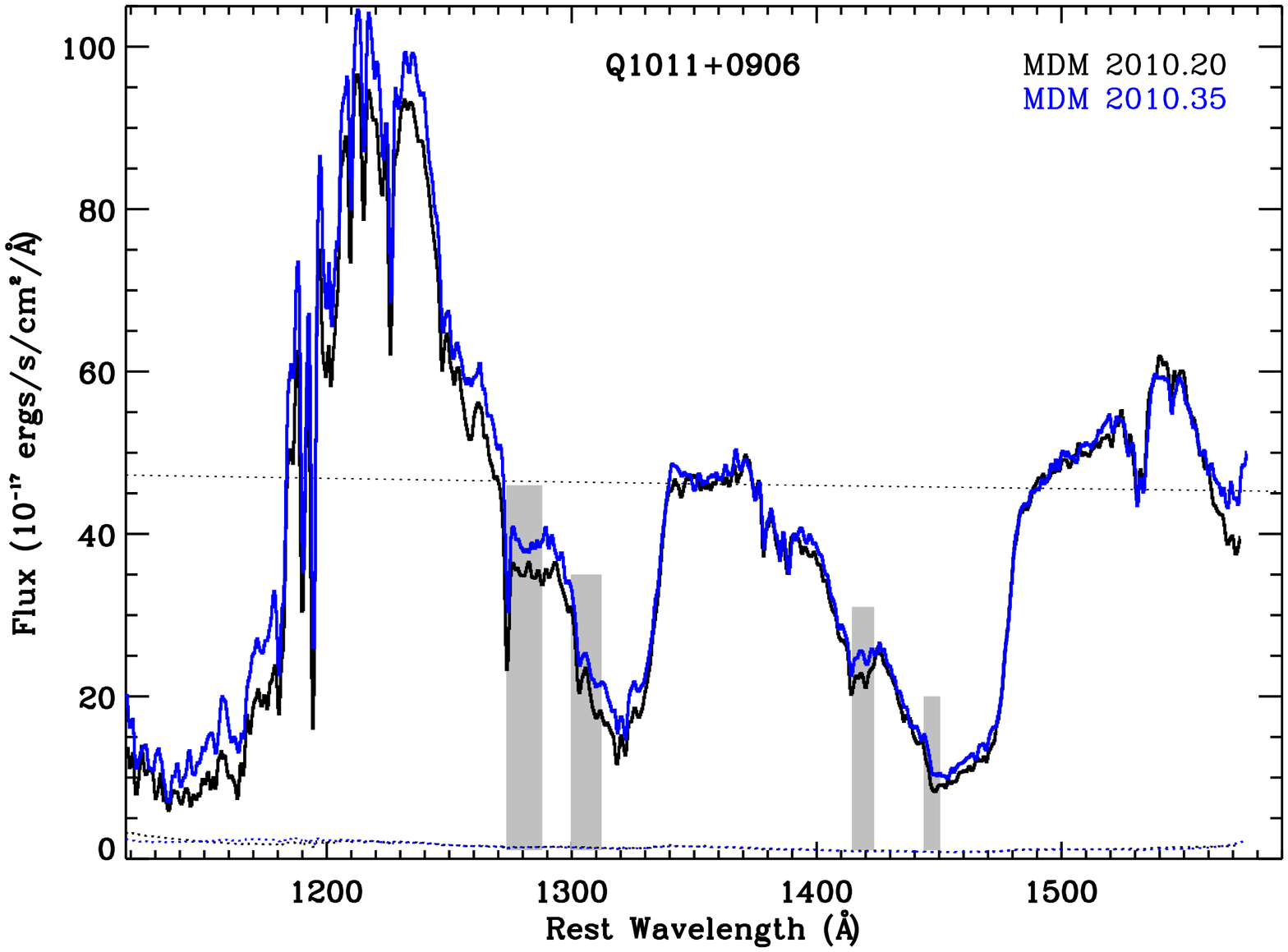}
  \caption{The full spectra of 1011$+$0906, showing 2 epochs covering a $\Delta$$t$ of 0.05 yr.
    We classify this measurement as a tentative case of variability. The rightmost shaded regions
    mark the candidate intervals of \civ\ variability, and the leftmost pair of shaded regions mark the
    corresponding velocities in \siiv.}
  \label{sp1011}
\end{figure*}

In this section, we describe a measurement that we classified as a tentative detection of variability. The full spectra for the 2010.20 and 2010.35 observations of 1011+0906, which are separated by a $\Delta$$t$ of 0.05 yr (17 days), are plotted in Fig. \ref{sp1011}, showing two potential regions of variability with shaded bars. These intervals are centered at $-$26 250 and $-$20 400 \kms\ and are 1900 and 1400 \kms\ wide, respectively. The significance of the variability is 9.1$\sigma$ in the wider interval and 6.9$\sigma$ in the narrower interval. These values are well above the threshold of 4$\sigma$, and there is potential variability in \siiv\ at the same velocities as in \civ. However, there appear to be mismatches throughout the spectrum, blueward of the \siiv\ emission line. These mismatches are similar in amplitude to the amplitude of the strength changes in the regions of potential \civ\ variability identified by the shaded regions in Fig. \ref{sp1011}. We considered the possibility that the slope of one of the spectra needs to be adjusted to better match the other. We examined the other data collected on the same nights as these two spectra, but found no evidence for any issues with the flux calibration at either epoch. However, we maintain our conservative approach here and keep this measurement in the ``tentative" category.

\subsubsection{0932+5006: Tentative variability over 31 days}
\label{ch4.0932}

\begin{figure}
  \includegraphics[width=84mm]{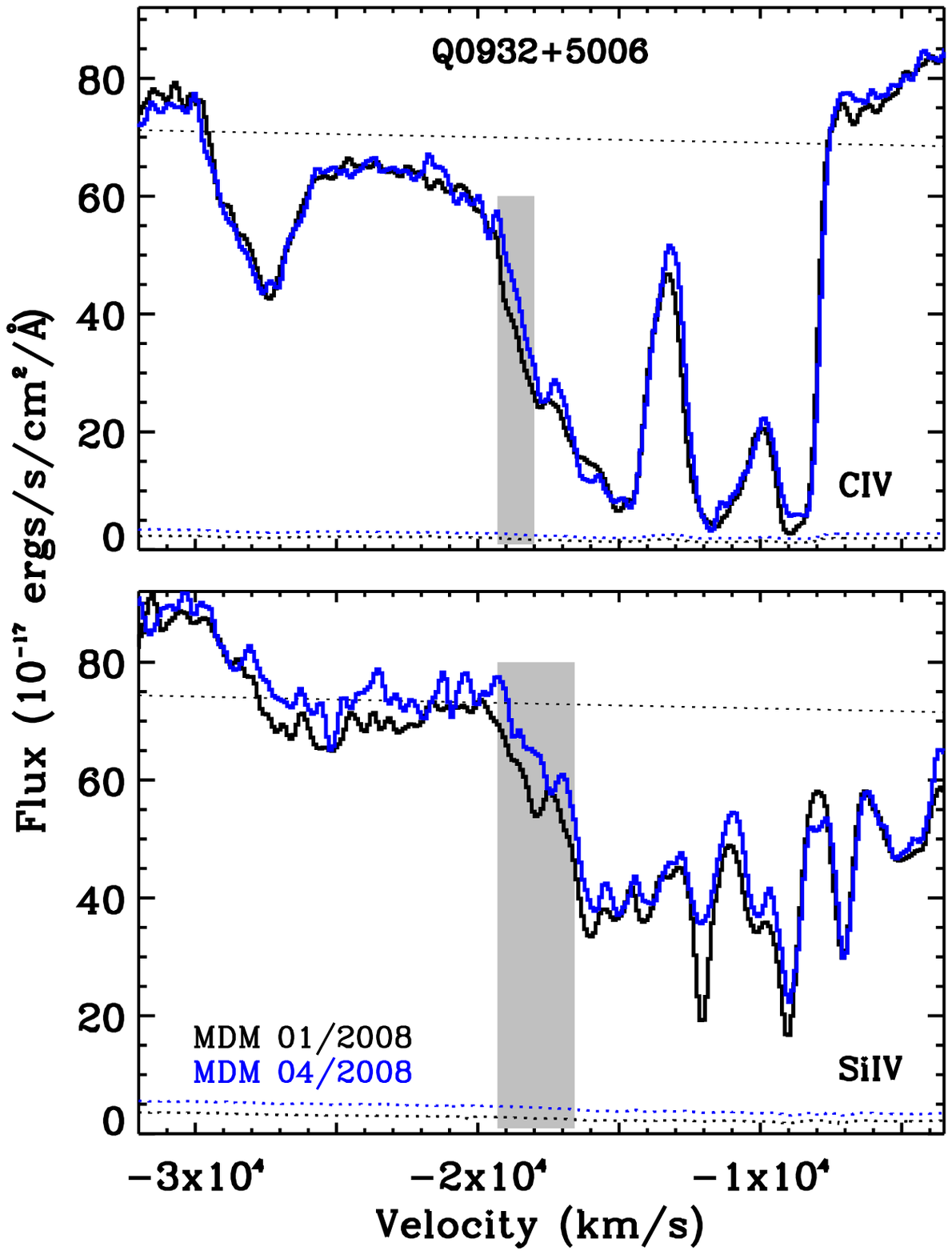}
  \caption{Spectra of the \civ\ (top panel) and \siiv\ (bottom panel) BALs in 0932$+$5006 for two
    epochs separated by 0.09 yr (31 days). This is another tentative measurement of variability.}
  \label{sp0932z}
\end{figure}

We present here another example of a tentative detection of variability. Fig. \ref{sp0932z} shows the \civ\ and \siiv\ BALs for the 2008.03 and 2008.28 observations of 0932+5006, which are separated by 0.09 yr (31 days). The shaded region identifies the candidate variable interval, with a width of 1200 \kms, which is just at the threshold defined in Paper 1. The flux difference within this interval is 8.3$\sigma$, and there is also possible variability at these velocities in \siiv. As in the case of 1011+0906, however, there are mismatches in other regions of the spectra, in particular towards the blue end of the spectra in the bottom panel of Fig. \ref{sp0932z}. Given the small amplitude of variability in an interval that is just wide enough to meet our threshold, combined with mismatches in other regions of the spectra, we classify this measurement as a tentative case of variability.

\subsubsection{0903+1734: Secure variability over 72 days}
\label{ch4.0903}

\begin{figure*}
  \includegraphics[width=100mm, angle=90]{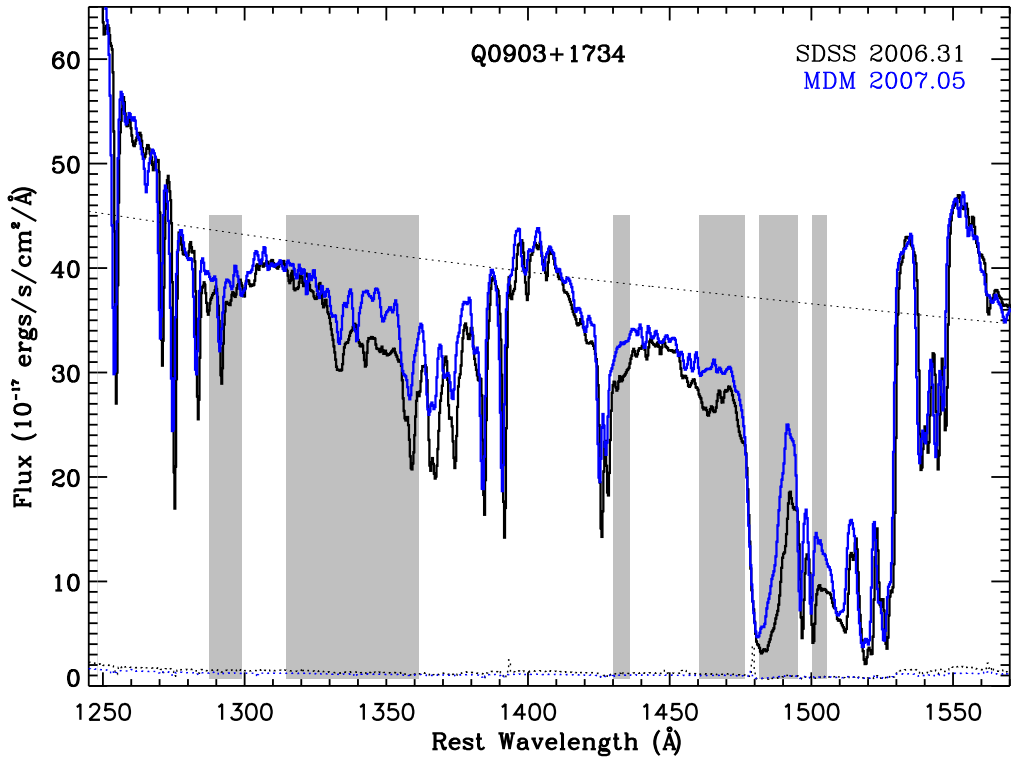}
  \caption{The full spectra of 0903+1734, showing 2 epochs with a $\Delta$$t$ of 0.20 yr. The four
    rightmost shaded regions mark the intervals of \civ\ variability, and the leftmost shaded regions
    mark the corresponding velocities in \siiv. Strong narrow lines located at $\sim$1425 \AA\ and
    blueward of 1300 \AA\ are due to ions at much lower redshifts, unrelated to the quasar.}
  \label{sp0903}
\end{figure*}

The variability between 2006.31 and 2007.05 in 0903+1734 is one of the most prominent cases of variability on time-scales less than those analyzed in Paper 1, i.e., $<$0.35 yr. These observations are separated by 0.20 yr (72 days), which is just on the upper threshold of the range of time-scales included in the fourth point of Fig. \ref{dream} below. Fig. \ref{sp0903} shows these two spectra with the variable intervals in \civ, and the corresponding velocities in \siiv, shaded. The variability here covers a wide range in velocities, unlike the other examples at shorter time-scales in this section. Paper 2 contains a full discussion on the multi-epoch behavior of this quasar.

\subsection{Time-scales for Variability}
\label{ch4.time}

\begin{figure}
  \includegraphics[width=84mm]{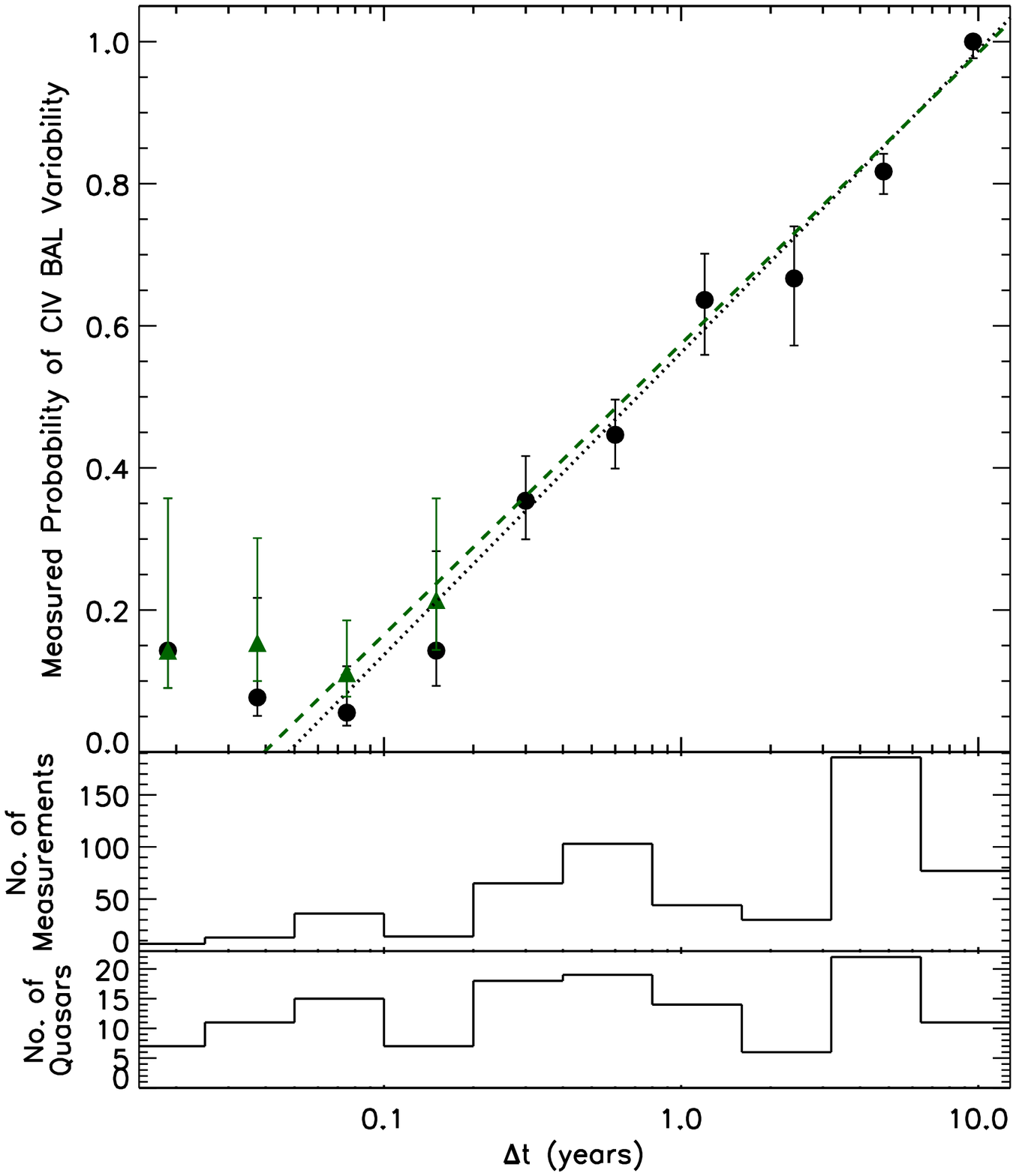}
  \caption{The top panel displays the fraction of measurements in which \civ\ BAL variability was
    detected at $>$4$\sigma$ over a velocity interval of at least 1200 \kms versus the time interval
    between the two observations. The green triangles show the variability fractions with tentative
    cases of variability included. The middle and bottom panels show the number of measurements
    and the number of quasars, respectively, that contribute to each bin.}
  \label{dream}
\end{figure}

In this section, we examine the relationship of \civ\ BAL variability with time-scale across the full measured range from 0.02 to 8.7 yr. To do this, we compare the BALs in each pair of observations at all velocities in every quasar and then count the occurrences of \civ\ BAL variability, using our definition of BAL variability first defined in Paper 1 (see Section \ref{ch4.meas}). We then calculate a probability by dividing the number of occurrences of variability by the number of measurements, where a pair of observations is one measurement, in logarithmic bins of $\Delta$$t$. We plot this measured probability of detecting \civ\ BAL variability versus $\Delta$$t$ in Fig. \ref{dream}. The 1$\sigma$ error bars are based on counting statistics for the number of occurrences of variability and the number of measurements in each bin \citep{Cameron11}.

Fig. \ref{dream} confirms our results from Paper 1 that the incidence of \civ\ BAL variability is higher when $\Delta$$t$ is larger. The probability of \civ\ variability approaches unity when $\Delta$$t$ $\geq$ 10 yrs. This decreases to $\sim$0.6 for $\Delta$$t$ $\sim$1 yr and $\sim$0.05$-$0.1 for $\Delta$$t$ $\sim$0.08 yr ($\sim$1 month). Below $\Delta$$t$ $\sim$ 0.08 yr, the measured probability of variability begins to increase slightly as $\Delta$$t$ becomes shorter. For the four shortest time-scale bins, where $\Delta$$t$ $<$0.2 yr, we also calculate this measured probability of variability with the tentative cases of variability, as described in Sections \ref{ch4.meas} and \ref{ch4.short}, included. The resulting values are plotted as green triangles in Fig. \ref{dream}. This tends to flatten out the trend in the shortest time-scale bins. The green and black dotted curves show the least-squares fit to the points with and without the tentative cases of variability included. Overall, inclusion of the tentative cases of variability does not significantly change the slope of the points. We note that these are probabilities for detecting variability between two measurements separated by $\Delta$$t$, and not for detecting variability at any time in that $\Delta$$t$ time-frame. In other words, if a quasar varied then returned to its initial state all within a certain $\Delta$$t$, then a measurement at that value of $\Delta$$t$ would not count as variable in this plot.

The values plotted in Fig. \ref{dream} represent probabilities only if each measurement is an independent event. There are two primary biases that affect this plot. First, some quasars contribute more than others (see Table 1), and those quasars that were observed most frequently throughout our monitoring programme were typically ones that were known to be variable (Section \ref{ch4.obs}). Second, our observations are clustered in time, with most taken at Lick from 1988 to 1992 and at MDM from 2007 to 2010. The middle and bottom panels of Fig. \ref{dream} display the number of measurements and the number of quasars, respectively, that contribute to each $\Delta$$t$ bin (each quasar can contribute multiple times to a single bin). The middle panel shows that the majority of the measurements are clustered at shorter ($<$1 yr) and longer ($>\sim$4 yr) time-scales. The average number of measurements per quasar in each bin ranges from 1.0 (in the shortest time-scale bin) to 8.5 (in the penultimate bin).
 
\begin{figure}
  \includegraphics[width=84mm]{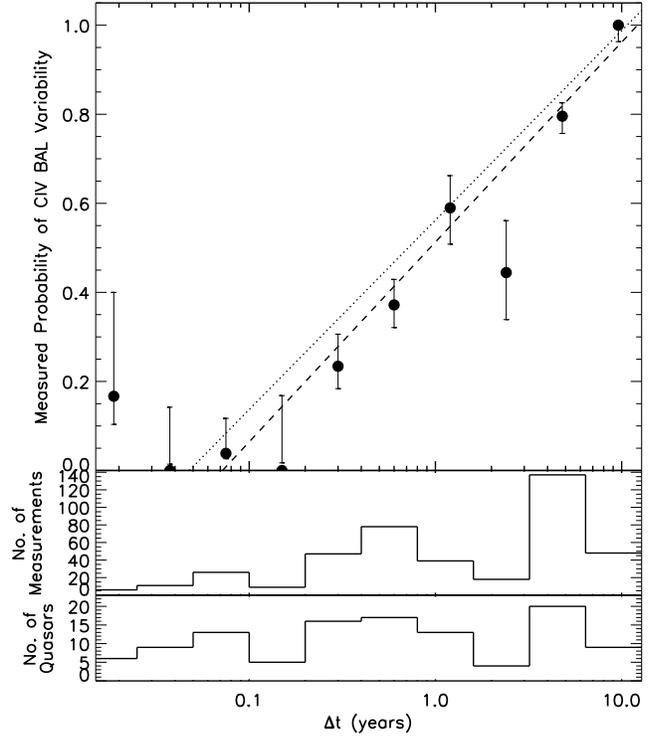}
  \caption{Same as Fig. \ref{dream}, but with the two most-observed quasars, 0842+3431 and
    0932+5006, removed. The dashed line in the top panel is a fit to the data points, and the fit from
    Fig. \ref{dream} is overplotted as a dotted line.}
  \label{dream_rem}
\end{figure}

Since those quasars contributing the most measurements to Fig. \ref{dream} are typically those that were known to be variable, the measured probabilities of variability in Fig. \ref{dream} may be biased to higher values. Those quasars with the most epochs of data will be counted much more often in Fig. \ref{dream}. For example, we have 13 epochs of data for 0842+3431 and 0932+5006, and up to just 10 epochs of data for each of the other quasars in the sample. Thirteen epochs for one quasar provides 78 pairs of observations, so 0842+3431 and 0932+5006 together contribute 27 per cent of the measurements in Fig. \ref{dream}. To address this bias, we therefore show in Fig. \ref{dream_rem} a version of this plot which omits these two quasars with the most data. We plot a least-squares fit to the data points in Fig \ref{dream_rem} (dashed curve) and overplot the least-squares fit from Fig. \ref{dream} (dotted curve) to compare their slopes. This shows that the most frequently observed quasars are not significantly affecting the slope of the points, but they do shift the line to slightly higher probabilities. We do not include the tentative cases of variability in this figure, or in Figs \ref{dream3} through \ref{dream_cum} below.

Even though the slope does not change significantly between Figs \ref{dream} and \ref{dream_rem}, there is one point at $\Delta$$t$ of 2.4 yr that does decrease significantly in Fig. \ref{dream_rem}. In Fig. \ref{dream}, there are only 6 quasars that contribute measurements to this bin because a spectrum from SDSS is necessary for having a measurement at these intermediate time-scales. We only have SDSS spectra for 8 out of the 24 quasars in our sample. By removing 0842+3431 and 0932+5006, we are left with just 4 quasars in this bin centered at 2.4 yr, and more than half of the measurements of variability in this bin in Fig. \ref{dream} are measurements of 0842+3431 and 0932+5006.

\begin{figure}
  \includegraphics[width=84mm]{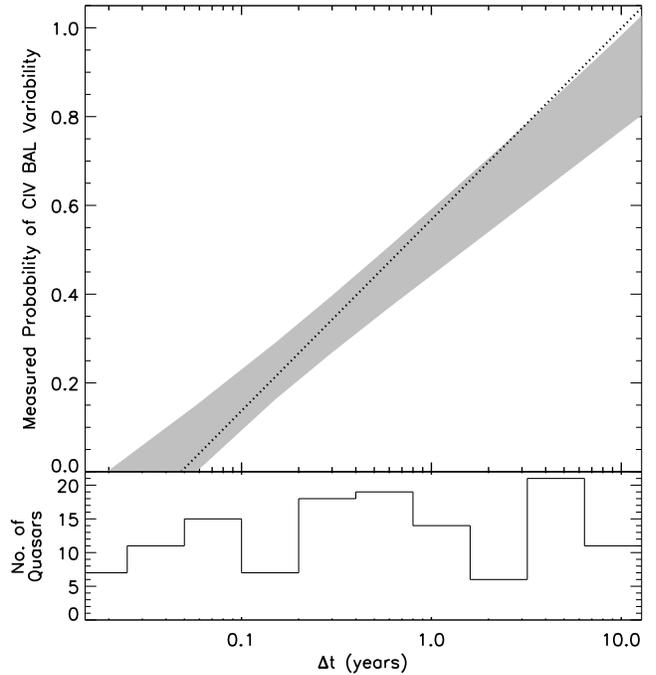}
  \caption{A version of Fig. \ref{dream} where each quasar only contributes once to each
    $\Delta$$t$ bin, and then we iterate 1000 times to get a range of least-square fits, shown here by
    the shaded region. The dotted line is the least-squares fit from Fig. \ref{dream}.}
  \label{dream3}
\end{figure}

This relates to the second bias mentioned above that instead of randomly sampling the quasar light curves, our measurements are clustered at shorter ($<$1 yr) and longer ($>$$\sim$4 yr) time-scales. For example, in cases where we have just Lick and MDM/KPNO data, any variability that we detect between the last Lick observation and the first MDM observation for that quasar could have occurred within any time interval during those $\sim$4 years. In this scenario, if there were, for example, 2 Lick observations and 5 MDM observations for an object, and the only variability that occurred in this quasar happened between the last Lick and the first MDM observations, then this quasar would contribute 10 measurements of variability to the last two $\Delta$$t$ bins. We therefore may be biased towards greater variability fractions at longer time intervals over intermediate and short time intervals.

To explore any bias that our uneven sampling may introduce into Fig. \ref{dream}, we created a version of this plot where each quasar can only contribute one measurement to each $\Delta$$t$ bin. If a quasar has multiple measurements in a particular bin, then one of those measurements is chosen at random to be included in this plot. We then recalculate this plot 1000 times, and for each iteration, we randomly choose which measurement per object is included in each bin and then create a least-squares fit to the resulting probability values. In Fig. \ref{dream3}, we show the range of least-squares fits we obtain. By iterating many times, we are attempting to uncover a range of slopes that includes what the slope would be in the ideal case of even sampling across the entire range of time-scales covered here. The range in slopes that we obtain overlaps the slope of the points in Fig. \ref{dream}, shown here again by the dotted curve, but the slopes tend to be shallower and/or shifted to slightly lower probabilities across the full range in $\Delta$$t$. This indicates that we do indeed have a slight bias towards greater variability fractions at longer time-scales in Fig. \ref{dream}. At the shortest time-scales, the curve from Fig. \ref{dream} is roughly in the middle of the range of curves in Fig. \ref{dream3}. We mention above in Section \ref{ch4.obs} that we tended to monitor those quasars known to vary, which might introduce a bias towards greater variability fractions at the shortest time-scales. However, it is not clear from Fig. \ref{dream3} whether such a bias exists, but if it does, it is certainly weaker than any bias at longer time-scales.

\begin{figure}
  \includegraphics[width=84mm]{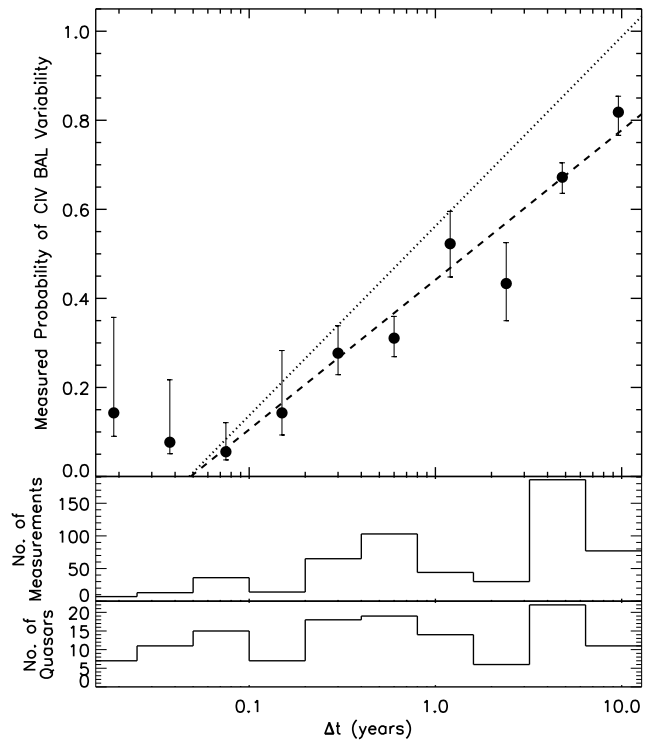}
  \caption{Same as Fig. \ref{dream}, but here we only consider absorption that lies between
    $-$25 000 and $-$3 000 \kms, thus satisfying all of the requirements of the balnicity index. The
    dotted line is the fit to the data points in Fig. \ref{dream}, and the dashed line is a fit to the data
    points in the current figure.}
  \label{dreamBI}
\end{figure}

Finally, we present Fig. \ref{dreamBI}, which shows the measured probability of \civ\ BAL variability versus time-scale, considering only the velocities that contribute to the measurement of BI, i.e., $-$25 000 to $-$3 000 \kms. Therefore, we only include absorption in this plot that are BALs according to the strict definition of \citet{Weymann91}, so the 2 quasars in our sample with BI=0, 0302+1705 and 0846+1540, are not included at all here. Restricting to this velocity range also removes uncertainty in the variability detection due to the presence of a potentially variable broad emission line (BEL).

The dotted line in Fig. \ref{dreamBI} represents the slope from Fig. \ref{dream} and the dashed line is a fit to the points in the current figure. These lines show that the restriction in velocity range changes the slope of the points; the slope becomes shallower. In Paper 1, we found that the incidence of variability increases at higher outflow speeds. Here, we are removing the highest-velocity measurements, so that will shift the points to lower values. If the incidence of variability at velocities $<$$-$25 000 \kms\ is the same across the entire range of time-scales, then the points at longer time-scales will be affected more than the points at smaller time-scales, where the overall incidence of variability is smaller. For example, if excluding the highest velocities lowers the recorded incidence of variability by 20 per cent at all time-scales, then the points at short-timescales, where the measured probability of CIV BAL variability is small, will only decrease very slightly, wherease the points at longer time-scales, where the probabilities are greater, would decrease by nearly 0.2. Thus, the slope in Fig. \ref{dreamBI} would naturally be shallower than in Fig. \ref{dream}.

\begin{figure}
  \includegraphics[width=84mm]{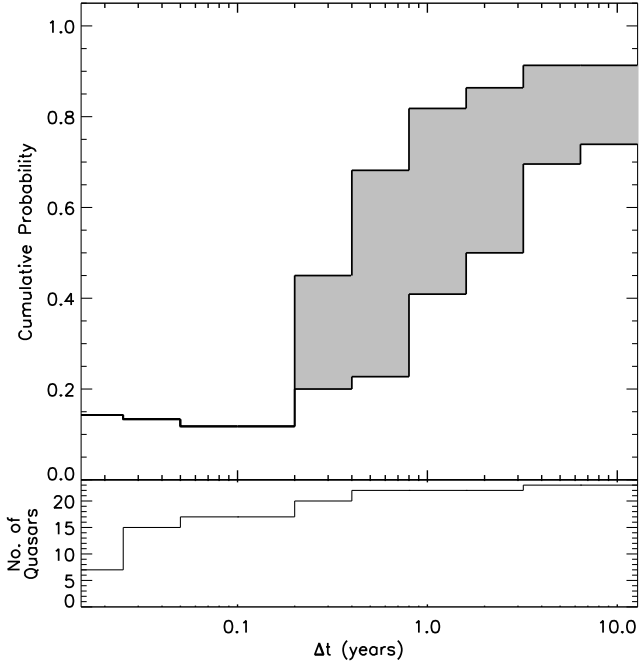}
  \caption{A cumulative distribution of the fraction of quasars with \civ\ variability with time-scale.
    This distribution is drawn from the full data set shown in Fig. \ref{dream}, excluding
    2225$-$0534. As in Fig. \ref{dream3}, each quasar only contributes once to each bin, and we
    iterate 1000 times to obtain a range of variability fractions at each time-scale. Because of the
    small number of measurements at the shortest time-scales, there is no range of values in the first
    four bins.}
  \label{dream_cum}
\end{figure}

We also investigate the cumulative probability of \civ\ BAL variability. In Fig. \ref{dream_cum}, each point represents the fraction of BAL \emph{quasars} with variable absorption at any $\Delta$$t$ value up to that point. We calculate this distribution using the full measurement sample in Fig. \ref{dream}, except that we omit 2225$-$0534 because we only have data over time-scales of $\Delta$$t$ $<$1 yr for this object. By the largest $\Delta$$t$ values, all the quasars in our sample, except for 2225$-$0534, are contributing. As in Fig. \ref{dream3}, we only allow each quasar to contribute once to each bin, and we then iterate 1000 times to obtain a range of variability fractions for each bin. This range is represented by the shaded region in Fig. \ref{dream_cum}, where the black histograms indicate the minimum and maximum variability fractions at each time-scale. If we included all measurements of each quasar in each bin, then as long as a quasar varied at least once in a particular bin, it would be included as variable in that bin and in all of the higher time-scale bins. In this case, the resulting curve would be the upper histogram in Fig. \ref{dream_cum}.

There is no range of values for the first four bins of Fig. \ref{dream_cum} because of the limited number of measurements per quasar in this time-scale regime. Below $\sim$0.2 yr, the fraction of quasars exhibiting \civ\ BAL variability is $\sim$13 per cent. At time-scales up to $\sim$1 yr, this fraction increases to 45 $\pm$ 23 per cent, and over multi-year time-scales, the variability fraction reaches 83 $\pm$ 9 per cent.

\section{Summary of Results}
\label{results}

This work is the third paper in a series on BAL variability in a sample of 24 BAL quasars, with observations covering a wide range of rest-frame time-scales from 0.02 to 8.7 yr. Paper 1 describes general trends in the \civ\ BAL variability data and finds, in particular, that variability occurs more often at higher outflow velocities and in shallower troughs (see also \citealt{Lundgren07,Gibson08,FilizAk12}). In both Paper 1 and the multi-epoch analysis of Paper 2, we note that variability typically occurs in just portions of troughs (e.g., Figs \ref{sp1246}$-$\ref{sp0842z}; see also \citealt{Gibson08}). In rare cases, BAL features appear, disappear, or change to or from narrower mini-BAL features (\citealt{Hamann08}; \citealt{Leighly09}; \citealt{Krongold10}; Papers 1 and 2; \citealt{Vivek12}; \citealt{FilizAk12}).

Paper 2 also directly compares variability in \civ\ absorption to variability in \siiv\ and determines that \siiv\ BALs are more likely to vary than \civ\ BALs. For example, at flow speeds $>$$-$20 000 \kms\ in the ``long-term" sample defined in Paper 1, 47 per cent of quasars exhibited \siiv\ variability while only 32 per cent exhibited \civ\ variability. Furthermore, approximately half of the variable \siiv\ regions did not have corresponding \civ\ variability at the same velocities, while in just one poorly measured case were changes in \civ\ not matched by variability in \siiv. The greater variability in \siiv\ is likely due to the tendency for weaker lines to vary more (Paper 1), combined with the fact that the \siiv\ absorption is typically weaker than \civ\ (Paper 2). Furthermore, when \civ\ and \siiv\ both varied at the same velocity, the changes always occured in the same sense (both lines either became stronger or both weakened). Similarly, BAL changes at different velocities within the same line almost always occurred in the same sense.

In the present paper, we examine \civ\ BAL variability behaviours as a function of time-scale, with a particular emphasis on the shortest times. Over 50 per cent of measurements separated by 1 yr in the quasar rest-frame varied, and on multi-year time-scales, the probability of detecting variability between two observations of the same quasar approaches unity (Fig. \ref{dream}). We therefore detect a strong trend towards greater variability fractions over longer time-scales, which is consistent with our findings in Paper 1.

Not only does the incidence of variability decrease at shorter time-scales, but the amplitude of the absorption strength changes ($\Delta$$A$) also decreases (Fig. \ref{dAvdt}). Over multi-year time-scales, we find a maximum $\Delta$$A = \,\sim$0.46. However, all of the variations across time-scales of $\Delta$$t$ $<$ 0.27 yr have amplitude changes of $\Delta$$A$ $<$ 0.2, and the shortest times we measure, $\Delta$$t$ $<$ 0.2 yr, have $\Delta$$A$ $<$ 0.1. This is generally consistent with \citet{Gibson08}, who find a decrease in $\Delta$EW values at shorter time-scales.

We detect variability down to the lowest $\Delta$$t$ values that we probe ($\Delta$$t$ $\sim$ 0.02 yr, or 8$-$10 days; Section \ref{ch4.short}). The amplitude changes on this short time-scale are small ($\Delta$$A$ $<$ 0.1), and they appear in just small portions of the BAL troughs (in intervals of $\Delta$$v$ from 1300 to 2600 \kms; Figs \ref{sp1246}$-$\ref{sp0842z} in Sections \ref{ch4.1246}$-$\ref{ch4.0842}).

We considered the possibility that the trend we find for greater variability fractions on longer time-scales may be biased due to uneven sampling of the quasar light curves and a tendency to monitor quasars that already varied. As discussed in Section \ref{ch4.time}, most of our observations are clustered at shorter ($<$1 yr) and longer ($>$$\sim$4 yr) time-scales. If we tend to monitor quasars already known to be variable, then those quasars will contribute many more measurements at $\Delta$$t$ $>$$\sim$4 yr than those that did not vary. However, Figs \ref{dream_rem} and \ref{dream3}, which remove the main sampling biases, still show strong trends for greater variability on longer time-scales. This trend is also readily apparent in Fig. \ref{dream_cum}, where we plot the fraction of quasars instead of the fraction of measurements. The fraction of quasars that exhibited variability at any $\Delta$$t$ $\le$$\sim$0.2 yr is $\sim$13 per cent. This fraction increases to 45 $\pm$ 23 per cent by $\Delta$$t$$\sim$1 yr and then to 83 $\pm$ 9 per cent over multi-year time-scales. This is greater than the value of 65 per cent we found for the two-epoch analysis on similar time-scales for this same sample of quasars in Paper 1. Although, we note that \citet{Gibson08} reported that 92 per cent (12/13) of their quasars exhibited \civ\ BAL variability in their two-epoch multi-year analysis.

\section{Discussion}
\label{discuss}

\subsection{Causes of Variability}
\label{causes}

The time-scales of quasar BAL variability are important for constraining the location of the outflowing gas, but the specific constraints are dependent on the cause(s) of the variability and the structure of the outflows. In Paper 2, we discuss two scenarios that could produce the observed BAL variability: 1) changes in the far-UV continuum flux causing global changes in the ionization of the outflowing gas, and 2) outflow clouds moving across our lines-of-sight to the quasar continuum source. We found evidence to support both scenarios, but for most individual cases, the cause of the variability is ambiguous.

In quasars that have variable absorption at more than one velocity, the variabilities tend to be coordinated (for example, 1246$-$0542, Fig. \ref{sp1246z}, and 0903+1734, Fig. \ref{sp0903}). \citet{Hamann11} found coordinated line variations in multiple narrow absorption line (NAL) systems and argue that this supports the scenario of a global change in ionization. Changes in the ionizing flux incident on the entire outflow should cause global changes in the ionization of the flow. A change in covering fraction due to moving clouds is less likely when absorption regions at different velocities vary in concert because this would require coordinated movements among outflow structures at different outflow velocities and radii.

On the other hand, variability in narrow portions of BAL troughs fits more naturally in the scenario of crossing clouds. This latter scenario is favored by previous work on BAL variability, including \citet{Lundgren07}, \citet{Gibson08}, \citet{Hamann08}, \citet{Krongold10}, \citet{Hall11}, and \citet{Vivek12}. \citet{Hamann08} and Capellupo et al, in preparation, discuss individual cases where the \civ\ BAL is saturated. Capellupo et al, in preparation, report on the detection of variable \pv\ \lam1118,1128 absorption at the same velocities as the \civ\ and \siiv\ absorption and variabilities in the same spectrum. The presence of a low-abundance line like \pv\ indicates that \civ\ is very saturated (\citealt{Hamann98}; \citealt{Hamann02}; Capellupo et al, in preparation; and references therein). The depths of saturated lines are governed by the covering fractions in those lines, so a change in covering fraction would change the depth of the line. A change in ionization would produce no detectable change in a highly saturated absorption trough. Therefore, variability in saturated absorption lines strongly favors the crossing cloud scenario. Furthermore, \citet{Moe09} detect a variable BAL component in a quasar spectrum, and they conclude that the variability was most likely caused by transverse motion of the outflow because they do not detect a significant change in the continuum flux of the quasar. However, variability on very short time-scales might pose a problem for the crossing cloud scenario, as we discuss below (Section \ref{clouds}).

Other possible causes for the BAL variability are more problematic. For example, a change in the size of the continuum source would cause a change in the covering fraction of saturated absorption lines. The main problem with this scenario, though, is that it should change the covering fraction of an entire trough and not just small portions thereof. Most cases of BAL variability occur in just portions of troughs (\citealt{Gibson08}; Papers 1 and 2).

Another possibility is instabilities within the flows themselves, as in the simulations of \citet{Proga00}, causing the observed variability. The time-scale for such instabilities significantly affecting the outflow structure is probably similar to the outflow dynamical time. If a flow has a radial velocity of 15 000 \kms\ and is located nominally around the broad-emission line region ($\sim$0.15 pc), the dynamical time is roughly $\sim$10 years. This is much larger than many of the measurements of variability time-scales in this work and cannot explain the very short time-scale changes we observe (as in Section \ref{ch4.short}). Furthermore, the time-scale for clouds dissipating without sheering or external forces is given by the sound crossing time. For a nominal cloud radius of 0.1 pc and temperature of 10$^{4}$ K, the sound crossing time is $\sim$8000 yr. This is much greater than the variability time-scales reported in this work, so dissipation of the outflow clouds is most likely not causing the observed variability.

As in Paper 2, we conclude here that, in general, the cause of variability is either changing ionization, transverse cloud movements, or some complex amalgam of the two. We therefore focus on just these two scenarios below.

\subsection{Implications of Variability}
\label{implicat}

\subsubsection{Changing ionization}
\label{ioniz}

Significant ionization changes in the BAL gas require at least a recombination time, nominally $t_{rec}\sim1/n_{e}\alpha_{r}$, if the gas maintains ionization equilibrium, where $n_{e}$ is the electron density and $\alpha_{r}$ is the recombination rate coefficient. For the cases of 1246$-$0542 and 0842+3431, we can use the variability time-scale of 8$-$10 days (Sections \ref{ch4.1246} and \ref{ch4.0842}) to set a lower limit on the density of $n_{e}$ $>$$\sim$4$-$5 x 10$^{5}$ cm$^{-3}$. Based on the photoionization calculations of \citet{Hamann95} and this constraint on the density from the recombination time, the maximum distance is $<$$\sim$200 pc, assuming the ionization is at least as high as that needed for a maximum \civ/C ratio. However, the actual recombination time can be shorter, resulting in lower derived densities and larger distances, if the gas is more highly ionized \citep{Krolik95,Hamann97}.

Another important implication of ionization changes is that they require significant variations in the quasars' incident (ionizing) flux. This seems unlikely for at least the shortest time-scales we measure because our sample consists of luminous quasars. More luminous quasars have a smaller amplitude of continuum variability than fainter AGN, and the amplitude of continuum variability tends to decrease on shorter time-scales (e.g. \citealt{VandenBerk04}). \citet{Misawa07} detected variability in a mini-BAL over a time-scale of 16 days, and they also state that variability on such a short time-scale is much faster than any expected changes in the continuum emission of a luminous quasar. They propose that a screen of gas, with varying optical depth, is located between the continuum source and the absorbing gas. This screen could be the ionized X-ray-shielding gas in the outflow models of \citet{Murray95}, and its proposed location is just above the accretion disk. The screen is in co-rotation with the disk, and if the screen is clumpy, then as it rotates, the intensity of the ionizing continuum that is transmitted through the screen varies. Since the screen is closer to the continuum source, it would be rotating at a faster velocity than the absorber and could then cause changes in the ionization of the absorbing gas more quickly than continuum variations alone. Several models indeed predict variability on very short time-scales related to this X-ray shielding gas (\citealt{Schurch09}; \citealt{Sim10}). \citealt{Schurch09} determine that the total column density in the gas could change by a factor of 2 in $\sim$9 days. (However, see \citealt{Hamann11} and in preparation for complications related to this scenario.) Another possibility for producing ionization changes on very short time-scales is small hot spots on an inhomogeneous accretion disk (e.g. \citealt{Dexter11}) which could appear and disappear on short time-scales. However, all of these scenarios predict global changes across BAL troughs, which we typically do not see (Section \ref{causes}).

\subsubsection{Crossing clouds}
\label{clouds}

We now consider a simplified scenario where the absorbers have a constant ionization and column density and the variability is due to a single outflow component moving across the continuum source. We can use the time-scale of the observed variability to estimate the crossing speed given a geometry for the emission and absorption regions. We estimate a characteristic diameter for the continuum source at 1500 \AA\ using the observed fluxes at this wavelength from \citet{Barlow93} and a standard bolometric correction factor, $L\approx 4.4\lambda L_{\lambda}(1500 {\rm \AA})$ \citep{Hamann11}, in a cosmology with $H_o = 71$ \kms\ Mpc, $\Omega_M = 0.3$, $\Omega_{\Lambda}=0.7$. This yields bolometric luminosities of $\sim2\times 10^{46}$ to $3\times 10^{47}$ ergs s$^{-1}$ across our sample. Based on the average bolometric luminosity, $\sim7\times 10^{46}$ ergs s$^{-1}$, a characteristic diameter for the continuum region at 1500 \AA\ is $D_{1500}\sim 0.008$ pc and for the \civ\ BEL region is $D_{\rm CIV}\sim 0.3$ pc, assuming $L = 1/3L_{edd}$ and $M_{BH} = 1.4\times10^{9}M_{\sun}$ (\citealt{Peterson04}; \citealt{Bentz07}; \citealt{Gaskell08}; Hamann \& Simon, in preparation).

\begin{figure}
  \centering
  \includegraphics[width=55mm]{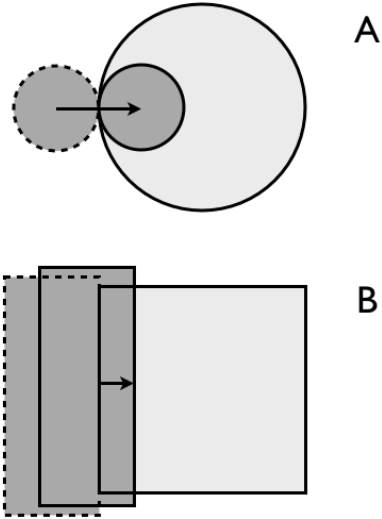}
  \caption{A simple schematic of the two models used to calculate crossing speeds of an outflow in
    the changing covering fraction scenario: (\emph{a}) the ``crossing disk" and (\emph{b}) the ``knife
    edge" model. The darker regions represent the outflow moving across the continuum source (light
    gray region) along our line-of-sight to the quasar. The arrows represent the distance the outflow
    must travel to cover the same fraction of the continuum source in both models \emph{a} and
    \emph{b}.}
  \label{schem}
\end{figure}

\begin{figure}
  \includegraphics[width=53mm, angle=90]{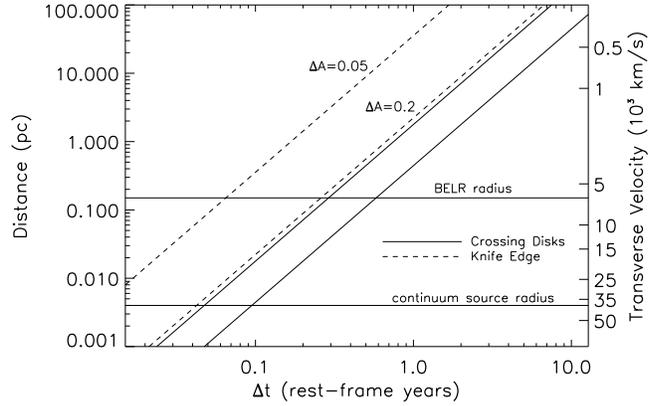}
  \caption{Transverse velocity and distance of the outflowing gas versus rest-frame time-scale for
    the two models illustrated in Fig. \ref{schem} and two typical values of $\Delta$$A$ (Fig.
    \ref{dAvdt}).}
  \label{dist}
\end{figure}

Using the characteristic diameter for the continuum region, we calculate the crossing speed of the outflow component for two models, as shown in Fig. \ref{schem}. In the first model, a circular disk crosses a larger circular continuum source along a path through the center of the background circle (the ``crossing disk" model). In the second model, the continuum source is square, and the moving absorber is larger, with a straight edge that moves across the background square (the ``knife edge" model). To calculate the crossing speed of the absorber, we take the change in strength of the line in the varying region, $\Delta$$A$, as the fraction of the continuum source that the absorber crosses within the elapsed time between observations, $\Delta$$t$. In the ``crossing disks" model, the distance traveled by the absorber is then $\sqrt{\Delta A}\,D_{1550}$, and the crossing speed is this distance traveled divided by $\Delta$$t$. This model gives the maximum transverse velocity. In the ``knife edge" model, the velocity is smaller than for the ``crossing disk" model by a factor of $\sqrt{\Delta A}$, and it gives the minimum transverse velocity. These two models represent opposite extremes which should encompass the possibilities for real absorbers. We plot the transverse velocity as a function of time-scale on the right ordinate in Fig. \ref{dist} for a change in BAL strength, $\Delta$$A$, of 0.05 and 0.2. A $\Delta$$A$ of 0.05 is a typical value for data on time-scales $<$0.1 yr, and values between $\Delta$$A$ of 0.05 and 0.2 are typical on time-scales of  0.1$-$3 yr (Fig. \ref{dAvdt}).

Next we tie the crossing speeds to a physical location by assuming they are roughly equal to the Keplerian rotation speed around an SMBH with a mass of $M_{BH} \sim 1.4\times10^{9}M_{\sun}$. This is a reasonable assumption for winds near their launch point from a rotating accretion disk (Section 1). However, beyond the launch radius, if there are only radial forces, the conservation of angular momentum will lead to a $\sim$1/r decline in the outflow transverse speeds. Magnetic fields threaded through the disk and applying a torque to the wind (e.g., \citealt{Everett05}) could counteract this decline. However, overall, we expect the actual transverse speeds to be less than or equal to the local Keplerian speed and, therefore, the distances derived from the Keplerian assumption are upper limits. (See below for more discussion.) These distances are shown as the left ordinate in Fig. \ref{dist}. This figure highlights the importance of obtaining measurements of BAL variability over shorter time-scales for constraining the location of the outflow. For example, variability times less than $\sim$0.2 yr (knife-edge) or 0.6 yr (crossing disk) indicate distances smaller than the nominal radius of the \civ\ broad emission line region (BELR).

The shortest time-scales over which we detect variability is 0.02$-$0.03 yr (8 and 10 days) in 1246$-$0542 and 0842+3431, with $\Delta$$A$ of 0.05$-$0.06. The transverse velocity of the outflow in these cases is 17 000 to 84 000 \kms\ for 1246$-$0542 and 18 000 to 71 000 \kms\ for 0842+3431, with the lower and upper limits given by the knife-edge and crossing disk models, respectively. These transverse velocities are surprisingly high, from $\sim$0.7 to 5 times the radial flow velocity, leading to derived distances of only $\sim$0.001 to 0.02 pc for both quasars, using the average SMBH mass for the quasar sample. For comparison, the estimated radius of the UV continuum source is 0.004 pc (see above). Similarly, \citet{Moe09} estimate a crossing speed of 18 000 \kms\ and a distance from the central SMBH $<$$\sim$0.1 pc, based on the variability in a single quasar.

These results provide much smaller distance constraints than the changing ionization scenario discussed above (Section \ref{ioniz}). There is indirect evidence for the crossing cloud scenario applying to 1246$-$0542 because its variations at $-$17 200 \kms\ occurred in what appears to be an optically thick trough (Figs \ref{sp1246} and \ref{sp1246z}). This interval lies in the bottom of the \civ\ BAL trough, at the same velocity as the bottom of the \siiv\ BAL trough. In Paper 2, we described how the \siiv\ optical depth should be $\sim$3.4 times less than \civ, given the lower abundance of Si compared to C when assuming solar abundances (\citealt{Hamann97}; \citealt{Hamann99}; \citealt{Asplund09}). Furthermore, if the ionization is at least as high as that needed for a maximum \civ /C ratio, then the optical depth in \siiv\ should be $>$8 times smaller than \civ\ \citep{Hamann11}. The apparent optical depth of \siiv\ in this interval is $\sim$0.5, indicating that the optical depth of \civ\ is at least $\sim$1.7$-$4. This is greater than the apparent \civ\ optical depth of $\sim$1.2, which suggests that the \civ\ line is saturated and would not likely be affected by modest changes in the ionization of the gas.

It is interesting to note that the crossing speeds implied by the shortest variability times are similar to or several times larger than the observed radial outflow speeds. In models of these flows driven by radiation pressure \citep{Murray95}, crossing speeds near or above the radial speed will occur only if the flow we measure is near its launch radius, e.g., where the gas is still roughly in corotation with the disk and the radial acceleration is not yet complete (also, \citealt{Murray98}). In models with significant magnetocentrifugal forces, there can be large components of both azimuthal and vertical velocity (perpendicular to the disk plane) far from the launch radius \citep{Everett05}. However, in luminous quasars with high accretion rates (i.e., high L/L$_{edd}$), radiative (and thus radial) forces are expected to dominate \citep{Everett07,Proga03a}. Therefore, the most likely explanation for the large crossing speeds in our study is that the measured flows are, indeed, at very small distances near their launch radius.

One final note on the distances that we calculate for BAL outflows is that previous studies have found evidence for absorbers located beyond the BEL region. \citet{Turnshek88} describe cases where a BAL overlaps a BEL in the spectra, and the absorption goes deeper than can be explained by the BAL absorbing just the continuum underneath the BEL. This indicates that, at least for these particular cases, the BAL region occults the BEL region. In our data, the \nv\ BAL in 0842+3431 might be deep enough to absorb both the continuum and some of the red side of the broad Ly$\alpha$ emission line. This would place the BAL gas farther out than the BEL region, at radii $r > 0.15$ pc, and apparently contradict our estimates above for $r\sim0.001-0.02$ pc based on the variability and crossing speeds. However, these assertions need further study because i) very few quasars show evidence for the BAL gas occulting the BELs, and ii) in the particular case of 0842+3431, the extension of the red wing in the Ly$\alpha$ emission profile is highly uncertain and the shape and depth of the observed \nv\ trough might be significantly contaminated by absorption in the Ly$\alpha$ forest.

\section*{Acknowledgments}

We thank an anonymous referee for helpful comments on the manuscript. We thank Paola Rodr\'iguez Hidalgo for helpful discussions. We acknowledge support from the National Science Foundation grant AST-0908910.

\bibliographystyle{mn2e}

\bibliography{full_bibliography}

\begin{thebibliography}{}

\bibitem[\protect\citeauthoryear{{Adelman-McCarthy et
  al.}}{{Adelman-McCarthy et al.}}{2008}]{Adelman08}
{Adelman-McCarthy} J.~K.,  {et al.} 2008, ApJS, 175, 297

\bibitem[\protect\citeauthoryear{{Asplund}, {Grevesse}, {Sauval} \&
  {Scott}}{{Asplund} et~al.}{2009}]{Asplund09}
{Asplund} M.,  {Grevesse} N.,  {Sauval} A.~J.,    {Scott} P.,  2009, ARA\&A,
  47, 481

\bibitem[\protect\citeauthoryear{{Barlow}}{{Barlow}}{1993}]{Barlow93}
{Barlow} T.~A.,  1993, PhD thesis, AA(California Univ.)

\bibitem[\protect\citeauthoryear{{Bentz}, {Denney}, {Peterson} \&
  {Pogge}}{{Bentz} et~al.}{2007}]{Bentz07}
{Bentz} M.~C.,  {Denney} K.~D.,  {Peterson} B.~M.,    {Pogge} R.~W.,  2007, in
  {Ho} L.~C.,  {Wang} J.-W.,  eds, The Central Engine of Active Galactic Nuclei
  Vol.~373 of Astronomical Society of the Pacific Conference Series, {Refining
  the Radius-Luminosity Relationship for Active Galactic Nuclei}.
p.~380

\bibitem[\protect\citeauthoryear{{Cameron}}{{Cameron}}{2011}]{Cameron11}
{Cameron} E.,  2011, \pasa, 28, 128

\bibitem[\protect\citeauthoryear{{Capellupo}, {Hamann}, {Shields},
  {Rodr{\'{\i}}guez Hidalgo} \& {Barlow}}{{Capellupo}
  et~al.}{2011}]{Capellupo11}
{Capellupo} D.~M.,  {Hamann} F.,  {Shields} J.~C.,  {Rodr{\'{\i}}guez Hidalgo}
  P.,    {Barlow} T.~A.,  2011, MNRAS, 413, 908

\bibitem[\protect\citeauthoryear{{Capellupo}, {Hamann}, {Shields},
  {Rodr{\'{\i}}guez Hidalgo} \& {Barlow}}{{Capellupo}
  et~al.}{2012}]{Capellupo12}
{Capellupo} D.~M.,  {Hamann} F.,  {Shields} J.~C.,  {Rodr{\'{\i}}guez Hidalgo}
  P.,    {Barlow} T.~A.,  2012, \mnras, 422, 3249

\bibitem[\protect\citeauthoryear{{Dexter} \& {Agol}}{{Dexter} \&
  {Agol}}{2011}]{Dexter11}
{Dexter} J.,  {Agol} E.,  2011, ApJL, 727, L24

\bibitem[\protect\citeauthoryear{{Di Matteo}, {Springel} \& {Hernquist}}{{Di
  Matteo} et~al.}{2005}]{DiMatteo05}
{Di Matteo} T.,  {Springel} V.,    {Hernquist} L.,  2005, Natur, 433, 604

\bibitem[\protect\citeauthoryear{{Everett}}{{Everett}}{2005}]{Everett05}
{Everett} J.~E.,  2005, ApJ, 631, 689

\bibitem[\protect\citeauthoryear{{Everett}}{{Everett}}{2007}]{Everett07}
{Everett} J.~E.,  2007, \apss, 311, 269

\bibitem[\protect\citeauthoryear{{Filiz Ak et al.}}{{Filiz Ak et
  al.}}{2012}]{FilizAk12}
{Filiz Ak} N.,  {et al.} 2012, \apj, 757, 114

\bibitem[\protect\citeauthoryear{{Gaskell}}{{Gaskell}}{2008}]{Gaskell08}
{Gaskell} C.~M.,  2008, in Revista Mexicana de Astronomia y Astrofisica
  Conference Series Vol.~32 of Revista Mexicana de Astronomia y Astrofisica
  Conference Series, {Accretion Disks and the Nature and Origin of AGN
  Continuum Variability}.
pp 1--11

\bibitem[\protect\citeauthoryear{{Gibson}, {Brandt}, {Gallagher}, {Hewett} \&
  {Schneider}}{{Gibson} et~al.}{2010}]{Gibson10}
{Gibson} R.~R.,  {Brandt} W.~N.,  {Gallagher} S.~C.,  {Hewett} P.~C.,
  {Schneider} D.~P.,  2010, ApJ, 713, 220

\bibitem[\protect\citeauthoryear{{Gibson}, {Brandt}, {Schneider} \&
  {Gallagher}}{{Gibson} et~al.}{2008}]{Gibson08}
{Gibson} R.~R.,  {Brandt} W.~N.,  {Schneider} D.~P.,    {Gallagher} S.~C.,
  2008, ApJ, 675, 985

\bibitem[\protect\citeauthoryear{{Hall}, {Anosov}, {White}, {Brandt}, {Gregg},
  {Gibson}, {Becker} \& {Schneider}}{{Hall} et~al.}{2011}]{Hall11}
{Hall} P.~B.,  {Anosov} K.,  {White} R.~L.,  {Brandt} W.~N.,  {Gregg} M.~D.,
  {Gibson} R.~R.,  {Becker} R.~H.,    {Schneider} D.~P.,  2011, MNRAS, 411,
  2653

\bibitem[\protect\citeauthoryear{{Hamann}}{{Hamann}}{1998}]{Hamann98}
{Hamann} F.,  1998, ApJ, 500, 798

\bibitem[\protect\citeauthoryear{{Hamann}, {Barlow}, {Beaver}, {Burbidge},
  {Cohen}, {Junkkarinen} \& {Lyons}}{{Hamann} et~al.}{1995}]{Hamann95}
{Hamann} F.,  {Barlow} T.~A.,  {Beaver} E.~A.,  {Burbidge} E.~M.,  {Cohen}
  R.~D.,  {Junkkarinen} V.,    {Lyons} R.,  1995, ApJ, 443, 606

\bibitem[\protect\citeauthoryear{{Hamann}, {Barlow}, {Junkkarinen} \&
  {Burbidge}}{{Hamann} et~al.}{1997}]{Hamann97}
{Hamann} F.,  {Barlow} T.~A.,  {Junkkarinen} V.,    {Burbidge} E.~M.,  1997,
  ApJ, 478, 80

\bibitem[\protect\citeauthoryear{{Hamann} \& {Ferland}}{{Hamann} \&
  {Ferland}}{1999}]{Hamann99}
{Hamann} F.,  {Ferland} G.,  1999, ARA\&A, 37, 487

\bibitem[\protect\citeauthoryear{{Hamann}, {Kanekar}, {Prochaska}, {Murphy},
  {Ellison}, {Malec}, {Milutinovic} \& {Ubachs}}{{Hamann}
  et~al.}{2011}]{Hamann11}
{Hamann} F.,  {Kanekar} N.,  {Prochaska} J.~X.,  {Murphy} M.~T.,  {Ellison} S.,
   {Malec} A.~L.,  {Milutinovic} N.,    {Ubachs} W.,  2011, MNRAS, 410, 1957

\bibitem[\protect\citeauthoryear{{Hamann}, {Kaplan}, {Rodr{\'{\i}}guez
  Hidalgo}, {Prochaska} \& {Herbert-Fort}}{{Hamann} et~al.}{2008}]{Hamann08}
{Hamann} F.,  {Kaplan} K.~F.,  {Rodr{\'{\i}}guez Hidalgo} P.,  {Prochaska}
  J.~X.,    {Herbert-Fort} S.,  2008, MNRAS, 391, L39

\bibitem[\protect\citeauthoryear{{Hamann}, {Sabra}, {Junkkarinen}, {Cohen} \&
  {Shields}}{{Hamann} et~al.}{2002}]{Hamann02}
{Hamann} F.,  {Sabra} B.,  {Junkkarinen} V.,  {Cohen} R.,    {Shields} G.,
  2002, in {Boller} T.,  {Komossa} S.,  {Kahn} S.,  {Kunieda} H.,   {Gallo} L.,
   eds, X-ray Spectroscopy of AGN with Chandra and XMM-Newton {How Massive are
  BALQSO Winds?}.
p.~121

\bibitem[\protect\citeauthoryear{{Krolik} \& {Kriss}}{{Krolik} \&
  {Kriss}}{1995}]{Krolik95}
{Krolik} J.~H.,  {Kriss} G.~A.,  1995, \apj, 447, 512

\bibitem[\protect\citeauthoryear{{Krongold}, {Binette} \&
  {Hern{\'a}ndez-Ibarra}}{{Krongold} et~al.}{2010}]{Krongold10}
{Krongold} Y.,  {Binette} L.,    {Hern{\'a}ndez-Ibarra} F.,  2010, ApJL, 724,
  L203

\bibitem[\protect\citeauthoryear{{Leighly}, {Hamann}, {Casebeer} \&
  {Grupe}}{{Leighly} et~al.}{2009}]{Leighly09}
{Leighly} K.~M.,  {Hamann} F.,  {Casebeer} D.~A.,    {Grupe} D.,  2009, ApJ,
  701, 176

\bibitem[\protect\citeauthoryear{{Lundgren}, {Wilhite}, {Brunner}, {Hall},
  {Schneider}, {York}, {Vanden Berk} \& {Brinkmann}}{{Lundgren}
  et~al.}{2007}]{Lundgren07}
{Lundgren} B.~F.,  {Wilhite} B.~C.,  {Brunner} R.~J.,  {Hall} P.~B.,
  {Schneider} D.~P.,  {York} D.~G.,  {Vanden Berk} D.~E.,    {Brinkmann} J.,
  2007, ApJ, 656, 73

\bibitem[\protect\citeauthoryear{{Misawa}, {Eracleous}, {Charlton} \&
  {Kashikawa}}{{Misawa} et~al.}{2007}]{Misawa07}
{Misawa} T.,  {Eracleous} M.,  {Charlton} J.~C.,    {Kashikawa} N.,  2007, ApJ,
  660, 152

\bibitem[\protect\citeauthoryear{{Moe}, {Arav}, {Bautista} \& {Korista}}{{Moe}
  et~al.}{2009}]{Moe09}
{Moe} M.,  {Arav} N.,  {Bautista} M.~A.,    {Korista} K.~T.,  2009, \apj, 706,
  525

\bibitem[\protect\citeauthoryear{{Moll}, {Schindler}, {Domainko}, {Kapferer},
  {Mair}, {van Kampen}, {Kronberger}, {Kimeswenger} \& {Ruffert}}{{Moll}
  et~al.}{2007}]{Moll07}
{Moll} R.,  {Schindler} S.,  {Domainko} W.,  {Kapferer} W.,  {Mair} M.,  {van
  Kampen} E.,  {Kronberger} T.,  {Kimeswenger} S.,    {Ruffert} M.,  2007,
  A\&A, 463, 513

\bibitem[\protect\citeauthoryear{{Murray} \& {Chiang}}{{Murray} \&
  {Chiang}}{1998}]{Murray98}
{Murray} N.,  {Chiang} J.,  1998, \apj, 494, 125

\bibitem[\protect\citeauthoryear{{Murray}, {Chiang}, {Grossman} \&
  {Voit}}{{Murray} et~al.}{1995}]{Murray95}
{Murray} N.,  {Chiang} J.,  {Grossman} S.~A.,    {Voit} G.~M.,  1995, ApJ, 451,
  498

\bibitem[\protect\citeauthoryear{{Peterson}, {Ferrarese}, {Gilbert}, {Kaspi},
  {Malkan}, {Maoz}, {Merritt}, {Netzer}, {Onken}, {Pogge}, {Vestergaard} \&
  {Wandel}}{{Peterson} et~al.}{2004}]{Peterson04}
{Peterson} B.~M.,  {Ferrarese} L.,  {Gilbert} K.~M.,  {Kaspi} S.,  {Malkan}
  M.~A.,  {Maoz} D.,  {Merritt} D.,  {Netzer} H.,  {Onken} C.~A.,  {Pogge}
  R.~W.,  {Vestergaard} M.,    {Wandel} A.,  2004, ApJ, 613, 682

\bibitem[\protect\citeauthoryear{{Proga}}{{Proga}}{2003}]{Proga03a}
{Proga} D.,  2003, \apj, 585, 406

\bibitem[\protect\citeauthoryear{{Proga}}{{Proga}}{2007}]{Proga07}
{Proga} D.,  2007, ApJ, 661, 693

\bibitem[\protect\citeauthoryear{{Proga} \& {Kallman}}{{Proga} \&
  {Kallman}}{2004}]{Proga04}
{Proga} D.,  {Kallman} T.~R.,  2004, ApJ, 616, 688

\bibitem[\protect\citeauthoryear{{Proga}, {Stone} \& {Kallman}}{{Proga}
  et~al.}{2000}]{Proga00}
{Proga} D.,  {Stone} J.~M.,    {Kallman} T.~R.,  2000, ApJ, 543, 686

\bibitem[\protect\citeauthoryear{{Schurch}, {Done} \& {Proga}}{{Schurch}
  et~al.}{2009}]{Schurch09}
{Schurch} N.~J.,  {Done} C.,    {Proga} D.,  2009, \apj, 694, 1

\bibitem[\protect\citeauthoryear{{Sim}, {Proga}, {Miller}, {Long} \&
  {Turner}}{{Sim} et~al.}{2010}]{Sim10}
{Sim} S.~A.,  {Proga} D.,  {Miller} L.,  {Long} K.~S.,    {Turner} T.~J.,
  2010, \mnras, 408, 1396

\bibitem[\protect\citeauthoryear{{Turnshek}}{{Turnshek}}{1988}]{Turnshek88}
{Turnshek} D.~A.,  1988, in {Blades} J.~C.,  {Turnshek} D.~A.,   {Norman}
  C.~A.,  eds, Proceedings of the QSO Absorption Line Meeting {BAL QSOs -
  Observations, models and implications for narrow absorption line systems}.
pp 17--46

\bibitem[\protect\citeauthoryear{{Vanden Berk}, {Wilhite}, {Kron}, {Anderson},
  {Brunner}, {Hall}, {Ivezi{\'c}}, {Richards}, {Schneider}, {York},
  {Brinkmann}, {Lamb}, {Nichol} \& {Schlegel}}{{Vanden Berk}
  et~al.}{2004}]{VandenBerk04}
{Vanden Berk} D.~E.,  {Wilhite} B.~C.,  {Kron} R.~G.,  {Anderson} S.~F.,
  {Brunner} R.~J.,  {Hall} P.~B.,  {Ivezi{\'c}} {\v Z}.,  {Richards} G.~T.,
  {Schneider} D.~P.,  {York} D.~G.,  {Brinkmann} J.~V.,  {Lamb} D.~Q.,
  {Nichol} R.~C.,    {Schlegel} D.~J.,  2004, ApJ, 601, 692

\bibitem[\protect\citeauthoryear{{Vivek}, {Srianand}, {Mahabal} \&
  {Kuriakose}}{{Vivek} et~al.}{2012}]{Vivek12}
{Vivek} M.,  {Srianand} R.,  {Mahabal} A.,    {Kuriakose} V.~C.,  2012, \mnras,
  421, L107

\bibitem[\protect\citeauthoryear{{Weymann}, {Morris}, {Foltz} \&
  {Hewett}}{{Weymann} et~al.}{1991}]{Weymann91}
{Weymann} R.~J.,  {Morris} S.~L.,  {Foltz} C.~B.,    {Hewett} P.~C.,  1991,
  ApJ, 373, 23

\end{thebibliography}

\bsp

\label{lastpage}

\end{document}